\documentclass[twocolumn,twocolappendix,table]{aastex7}

\usepackage{CJK}
\usepackage{amsmath}
\usepackage{booktabs}
\usepackage{mathrsfs}
\usepackage{lipsum}
\usepackage{hyperref}

\accepted{March 1, 2026}

\submitjournal{APJ}

\begin{document}
 \begin{CJK*}{UTF8}{gbsn}
 
\title{Exploring the $S_8$ Tension: Insights from the CatNorth 1.5-Million Quasar Candidates}


\correspondingauthor{Jin Qin, Xue-Bing Wu}

\author[0000-0002-7533-1264]{Jin Qin (秦晋)}
\affiliation{Department of Astronomy, School of Physics, Peking University, Beijing 100871, People's Republic of China}
\affiliation{Kavli Institute for Astronomy and Astrophysics, Peking University, Beijing 100871, People's Republic of China}
\email[show]{qinjin@stu.pku.edu.cn}

\author[0000-0002-7350-6913]{Xue-Bing Wu (吴学兵)}
\affiliation{Department of Astronomy, School of Physics, Peking University, Beijing 100871, People's Republic of China}
\affiliation{Kavli Institute for Astronomy and Astrophysics, Peking University, Beijing 100871, People's Republic of China}
\email[show]{wuxb@pku.edu.cn}

\author[0000-0002-0759-0504]{Yuming Fu (傅煜铭)}
\email{yfu@strw.leidenuniv.nl}
\affiliation{Leiden Observatory, Leiden University, Einsteinweg 55, 2333 CC Leiden, The Netherlands}
\affiliation{Kapteyn Astronomical Institute, University of Groningen, P.O. Box 800, 9700 AV Groningen, The Netherlands}

\author[0000-0003-1132-8258]{Haojie Xu (许浩杰)}
\affiliation{Shanghai Astronomical Observatory, Chinese Academy of Sciences, Nandan Road 80, Shanghai 200240, China}
\affiliation{Department of Astronomy, Shanghai Jiao Tong University, Shanghai 200240, China}
\affiliation{Key Laboratory for Particle Astrophysics and Cosmology (MOE)/Shanghai Key Laboratory for Particle Physics and Cosmology, Shanghai 200240, China}
\email{haojie.xu@shao.ac.cn}

\author[0009-0005-3823-9302]{Yuxuan Pang (庞宇萱)}
\affiliation{Department of Astronomy, School of Physics, Peking University, Beijing 100871, People's Republic of China}
\affiliation{Kavli Institute for Astronomy and Astrophysics, Peking University, Beijing 100871, People's Republic of China}
\email{1901110223@pku.edu.cn}

\author[0000-0003-4916-6346]{Yun-Hao Zhang (张云皓)}
\affiliation{Leiden Observatory, Leiden University, Einsteinweg 55, 2333 CC Leiden, The Netherlands}
\affiliation{Institute for Astronomy, University of Edinburgh, Royal Observatory, Blackford Hill, Edinburgh, EH9 3HJ, The United Kingdom}
\email{YunHao.Zhang@ed.ac.uk}

\author[0000-0003-2632-9915]{Pengjie Zhang (张鹏杰)}
\affiliation{Department of Astronomy, Shanghai Jiao Tong University, Shanghai 200240, China}
\affiliation{Key Laboratory for Particle Astrophysics and Cosmology (MOE)/Shanghai Key Laboratory for Particle Physics and Cosmology, Shanghai 200240, China}
\affiliation{Tsung-Dao Lee Institute, Shanghai Jiao Tong University, Shanghai 200240, China}
\email{zhangpj@sjtu.edu.cn}

\begin{abstract}
The parameter $S_8$, a key probe of cosmic structure growth, exhibits a persistent $\sim3\sigma$ tension between high-redshift measurements from cosmic microwave background (CMB) anisotropies and low-redshift weak gravitational lensing observations. This discrepancy may indicate either unaccounted systematic effects or new physics beyond the standard $\Lambda$CDM cosmology. In this work, we constrain $S_8$ using the high purity CatNorth 1.5 million quasar candidate catalog and the {\it Planck} DR4 CMB lensing data across the broad redshift ranges through auto-correlation and cross-correlation analyses. To address the spatial incompleteness, we develop a machine-learning-based selection function that effectively suppresses the systematics-induced power spectrum excess on large scales. Our robust low-redshift measurements at $z<1.5$ yield $S_8 = 0.844^{+0.058}_{-0.056}$, consistent with the {\it Planck} 2018 CMB anisotropies constraints of $S_8=0.834\pm0.016$ but lower than the $0.879^{+0.055}_{-0.055}$ reported by a previous work using the Quaia quasar candidate catalog. However, for high-redshift faint quasars at $z>1.5$, we find a lower value of $S_8=0.724^{+0.058}_{-0.054}$, likely due to the sample incompleteness and/or the foreground contamination. Further tests on the volume-limited samples exhibit a consistent trend: $S_8 = 0.835^{+0.053}_{-0.049}$ for $z < 2$, $0.824^{+0.061}_{-0.062}$ for $0.4 < z < 1.5$, and a lower value of $0.789^{+0.062}_{-0.062}$ for the higher redshift range of $1.5 < z < 2.5$. While future data may refine these results, our current measurements based on a large sample of quasar candidates show less evidence of the $S_8$ tension.
\end{abstract}

\section{Introduction} 
\label{sec:intro}
Since the Type Ia supernova observations confirmed the accelerated expansion of the universe \citep{1998_Riess_supernova,1999_Perlmutter_supernova}, the $\Lambda$ cold dark matter ($\Lambda$CDM) model, in which dark energy is interpreted as the cosmological constant $\Lambda$, has become the standard cosmological model over the past two decades. The primary measurements to verify the standard model come from the cosmic microwave background (CMB) anisotropies \citep{2018_planck_cosmology_parameters} and large-scale structure (LSS) tracers \citep{2021_eBoss,2021_kids1000_cosmic_shear,2021_kids1000_weak_lensing_galaxy_clustering}. Despite the increasing precision of CMB and LSS measurements supporting the $\Lambda$CDM model, several observational tensions have emerged, challenging the model's validity. Among these, one of the most prominent tensions is the $S_8$ tension, which refers to the discrepancy between different measurements of cosmic structure. The {\it Planck}-2018 CMB anisotropy data, assuming a $\Lambda$CDM model, provides the best-fit value of $S_8 = 0.834 \pm 0.016$ \citep{2018_planck_cosmology_parameters}. However, this result is in $2\sim3\,\sigma$ tension with the cosmic shear measurements from the weak gravitational lensing (WGL) surveys. The KiDS-1000 analysis yields $S_8 = 0.759^{+0.024}_{-0.021}$ \citep{2021_kids1000_cosmic_shear}, corresponding to a $3\,\sigma$ discrepancy with {\it Planck}. The HSC-Y3 analysis reports $S_8 = 0.776^{+0.032}_{-0.033}$ \citep{2023_HSC_cosmology}, indicating a $2\,\sigma$ tension. The DES-Y3 analysis finds $S_8 = 0.759^{+0.025}_{-0.023}$ \citep{2022_DESY3_cosmic_shear_cosmology}, with a $2.3\,\sigma$ deviation. A more precise $3 \times 2$pt analysis of DES-Y3 data, which combines galaxy clustering and weak lensing, provides a refined constraint of $S_8 = 0.776 \pm 0.017$ \citep{2022_DESY3_3_cross_2_cosmology}, still showing a mild tension with {\it Planck}. Redshift Space Distortion (RSD) data also exhibits a $3.1\,\sigma$ discrepancy with {\it Planck} \citep{2018_Kazantzidis_S8_from_RSD,2021_Nunes_S8_from_RSD}. These tensions may point to unaccounted physical mechanisms \citep{2021_Lu_impact_of_baryons} or unknown systematic errors in the analysis of cosmic shear data \citep{2022_weak_lensing_systematics,2024_weak_lensing_systematics}. 
However, not all analyses show a significant discrepancy. Several studies report no such tension. For example, the galaxy cluster abundance results of \citet{1997_FanXiaohui_S8_cluster} give $\sigma_8 = 0.85 \pm 0.15$ for $\Omega_m = 0.3 \pm 0.1$, while the RSD measurements with SDSS DR7 groups and clusters \citep{2012_ChengLi_S8} yield $\sigma_8 = 0.86 \pm 0.03$, corresponding to $S_8 \sim 0.82$. In addition, the recent eROSITA results report $S_8 = 0.86 \pm 0.01$ \citep{2024_eROSITA_S8}, and the KiDS-Legacy survey finds $S_8 = 0.815^{+0.016}_{-0.021}$ \citep{2025_kids_legacy_cosmologicalconstraintscosmic}, which is consistent with the {\it Planck} 2018 constraints within 1.5$\sigma$. Therefore, the current situation remains inconclusive. To better understand the origin of these discrepancies, it is essential to obtain complementary constraints on $S_8$ that are independent of both CMB anisotropies and galaxy cosmic shear.

Quasars, as the most luminous active galactic nuclei (AGN), are versatile cosmological probes due to their high luminosities and wide redshift coverage. Their applications in cosmology span several key areas. First, quasars provide valuable insights into the expansion history of the Universe. Their spatial distribution enables measurements of baryon acoustic oscillations (BAO), which serve as standard rulers to constrain the Hubble parameter $H(z)$ \citep{2025_Adame_DESI_III,Adame_2025_DESI_VI}. Quasars observed in strong gravitational lensing systems can be used for time-delay cosmography, offering an independent method to determine the Hubble constant $H_0$ \citep{2019_wong_QSO_time_delay}. In recent years, quasars have also been explored as standard candles through empirical correlations between their X-ray and ultraviolet fluxes, enabling the construction of a quasar-based Hubble diagram to probe the expansion history and constrain the cosmological constant parameter $\Omega_\Lambda$ \citep{2015_Risaliti_quasar_Hubble_diagram,2019_Risaliti_quasar_hubble_diagram}. Furthermore, quasar spectra allow for the implementation of the Sandage test (redshift drift), a method entirely independent of standard rulers and candles, capable of directly measuring $H(z)$ \citep{1962_sandage_test,2008_sandage_test,2023_sandage_test}. Second, quasars are excellent tracers of cosmic structure formation. Their spatial clustering reflects the underlying matter density field and its evolution \citep{2012_BOSS}. Redshift-space distortions (RSD) in quasar clustering enable the measurements of the structure growth rate, quantified by $f\sigma_8$ \citep{2020_SDSS_IV_BAO_RSD}, thereby providing further constraints on the gravitational clustering and the nature of gravity. In addition, the absorption features in quasar spectra—such as the Lyman-alpha forest and the Gunn-Peterson trough—trace the distribution of neutral hydrogen, offering insights into the epoch of reionization \citep{2006_fan_cosmic_reionization,2023_fan_cosmic_reionization}.

Quasars offer valuable insights into the $S_8$ tension by combining their auto-correlation with cross-correlation against CMB lensing maps. Since its first detection, the quasar-CMB lensing cross-correlation has received growing attention. It has been used to probe the growth of cosmic structures up to redshift $z \sim 2$ \citep{2021_Garcia-Garcia_quasar_growth_history} and to constrain the primordial non-Gaussianity \citep{2024_Krolewski_PNG}, thereby contributing to the tests of inflationary models. Recently, \cite{2023_Alonso_quaia_structure_growth} employed Quaia, a Gaia- and unWISE-based catalog of over one million quasar candidates \citep{2024_Storey-Fisher_quaia_catalog}, along with the {\it Planck} DR4 CMB lensing data \citep{2022_Carron_CMB_lensing} to constrain the growth of structure. They divided the Quaia sample into low- and high-redshift subsamples, yielding $S_8 = 0.879 \pm 0.055$ and $S_8 = 0.729 \pm 0.063$, respectively. They attributed the lower $S_8$ value in the high-redshift sample to the contamination of the CMB lensing map by high-redshift extragalactic foregrounds. A joint analysis of both subsamples yielded $S_8 = 0.819 \pm 0.042$, consistent with {\it Planck} at the $1.4\, \sigma$ level.

However, current cross-correlation studies involving quasars and CMB lensing face several challenges. First, identifying quasars is inherently more complex than selecting galaxies, and the traditional color-color selection methods require further refinement \citep{Peters_2015_Quasar_classification_using_color_and_variability}. Second, the photometric redshift estimates for quasars are complicated by strong emission lines, which shift in and out of broadband filters at different redshift, causing color degeneracies \citep{2017_Yang_Quasar_Photometric_Redshifts_and_Candidate_Selection} and increasing uncertainties. These redshift errors propagate into the cosmological parameter constraints. In addition, quasar clustering measurements are particularly vulnerable to biases from spatial depth variations and stellar contamination. If uncorrected, these systematics can significantly affect the cosmological conclusions on the largest scales \citep{2013_pullen_systematic_effects,2013_Leistedt_systematics}.

In this work, we address these challenges by combining the novel CatNorth quasar candidate catalog \citep{Fu_2024_CatNorth} with the {\it Planck} CMB lensing data \citep{2022_Carron_CMB_lensing}, leveraging machine learning to mitigate systematics and obtain robust constraints on cosmological parameters, particularly $S_8$. The CatNorth catalog contains 1.5 million quasar candidates with high purity and accurate photometric redshifts, significantly alleviating the challenges discussed above and making it well-suited for cosmological analyses. To address the spatial incompleteness of CatNorth quasar candidates, we develop a machine learning-based selection function trained on systematics templates. This method effectively suppresses the power spectrum excesses induced by systematics. Compared to the Gaussian Process approach used in \cite{2024_Storey-Fisher_quaia_catalog} for the Quaia catalog, our method is faster, more memory-efficient, and readily extendable to broader cosmological applications.

\begin{figure}
    \centering
    \includegraphics[width=1.0\linewidth]{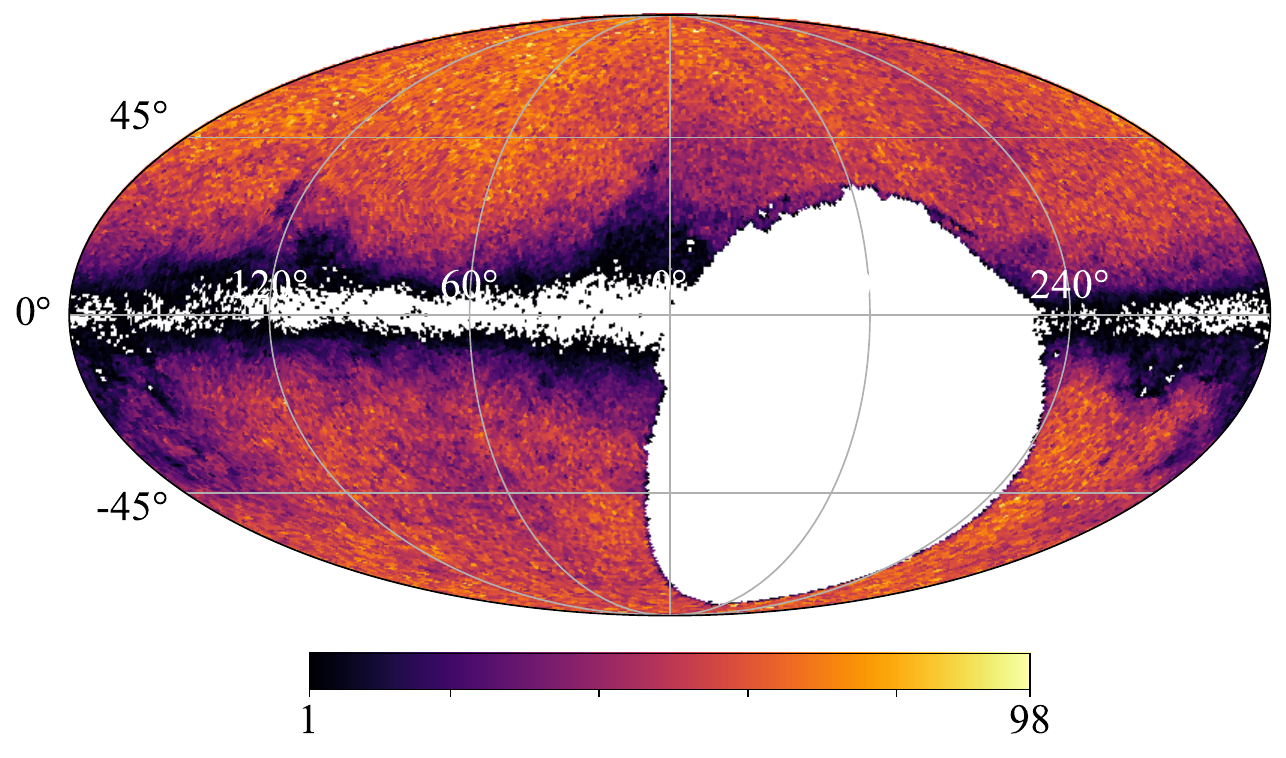}
    \caption{The sky density map of the CatNorth quasar candidate catalog \citep{Fu_2024_CatNorth} is shown in Galactic coordinates using the \texttt{HEALPix} framework \citep{2005_Gorski_healpix}. The map adopts a resolution parameter of $N_{\rm side} = 64$, corresponding to an area of 0.839 square degrees per pixel. Accordingly, the color bar is expressed in units of counts per 0.839 square degrees.}
    \label{fig:catnorth}
\end{figure}

\begin{figure}
    \centering    
    \includegraphics[width=1.0\linewidth]{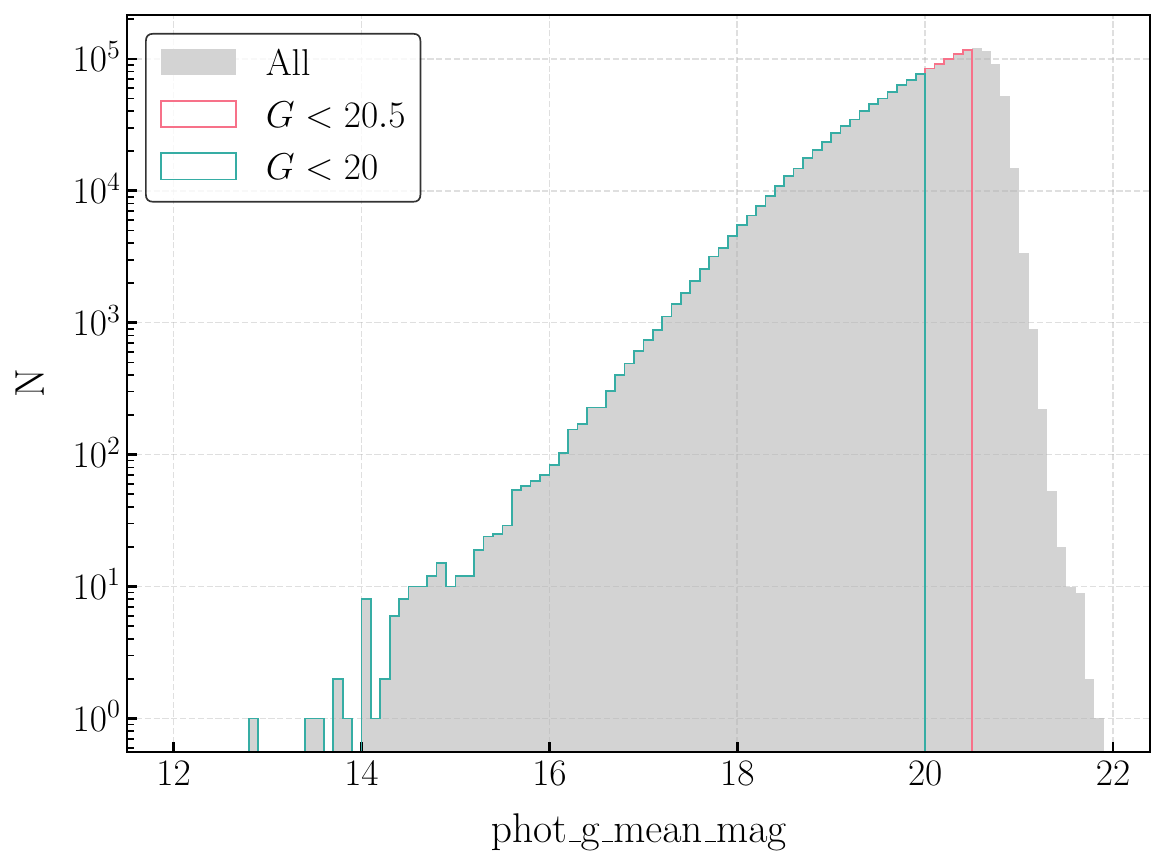}
    \caption{The apparent magnitude distribution of the CatNorth objects, where we mark the Gaia G-band magnitude limits of 
\(\texttt{phot\_g\_mean\_mag} \leq 20.0\) and \(\texttt{phot\_g\_mean\_mag} \leq 20.5\).}
    \label{fig:mag dist}
\end{figure}

The paper is structured as follows. Section~\ref{sec:data} introduces the CatNorth quasar candidate catalog and the {\it Planck} CMB lensing map used in our analysis. Section~\ref{sec:meth} outlines the theoretical framework for constructing the selection function and conducting the angular power spectrum analysis. Section~\ref{sec:results} presents the results and discusses their implications. We give our conclusions in Section~\ref{sec:conclusion}. Additional technical details are provided in the Appendix.

\section{Data} \label{sec:data}

\subsection{The CatNorth Quasar Candidate Catalog}

The CatNorth quasar candidate catalog contains over 1.5 million sources spanning the $3\pi$ sky \citep[see Figure \ref{fig:catnorth}]{Fu_2024_CatNorth}. Although it includes a small number of sources at declinations below $-30^\circ$, we exclude them from our analysis to simplify the processing. CatNorth is constructed by combining the Gaia Data Release 3 (hereafter Gaia DR3;\citealt{GaiaDataRelease3}) quasar candidate catalog with optical photometry from Pan-STARRS1 (five bands: \texttt{g, r, i, z, y}) and infrared photometry from CatWISE2020 (two bands: \texttt{W1, W2}). The magnitude distribution of CatNorth sources is shown in Figure \ref{fig:mag dist}, demonstrating that the catalog is nearly complete down to $\texttt{phot\_g\_mean\_mag} \leq 20.5$.

To address the key challenges in the current cross-correlation studies involving quasars and CMB lensing, CatNorth adopts a machine learning-based classification strategy. Specifically, an XGBoost classifier is used to improve upon the traditional color-color selection methods. This approach significantly increases the purity of the Gaia DR3 quasar candidate sample from 52\% to approximately 90\%, while maintaining high completeness. With a sky coverage of $3\pi$ steradians and a magnitude limit of $G<20.5$, CatNorth contains 1,148,821 sources—more than Quaia, which includes 1,020,271 sources over the same sky area and at the same magnitude limit. Furthermore, CatNorth exhibits notably higher completeness near the faint-end limit ($G = 20.5$), while preserving a comparable level of quasar purity \citep{Fu_2024_CatNorth}.

In addition to the improved classification, CatNorth provides enhanced redshift estimates. Photometric redshifts for all candidates are inferred using an ensemble regression model, and spectroscopic redshifts for 89,100 sources are derived from Gaia BP/RP spectra using a convolutional neural network. These improvements significantly reduce the emission-line misidentifications—such as the confusion between the C IV and C III] lines—and improve the overall redshift accuracy. In comparison, while Quaia represents a substantial improvement over Gaia DR3, it still inherits certain line misidentifications, such as the frequent confusion of the C IV line with Ly$\alpha$. Together, these enhancements make CatNorth a more robust and precise quasar candidate catalog for cosmological applications \citep{Fu_2024_CatNorth}.

\begin{figure}
    \centering
    \includegraphics[width=1.0\linewidth]{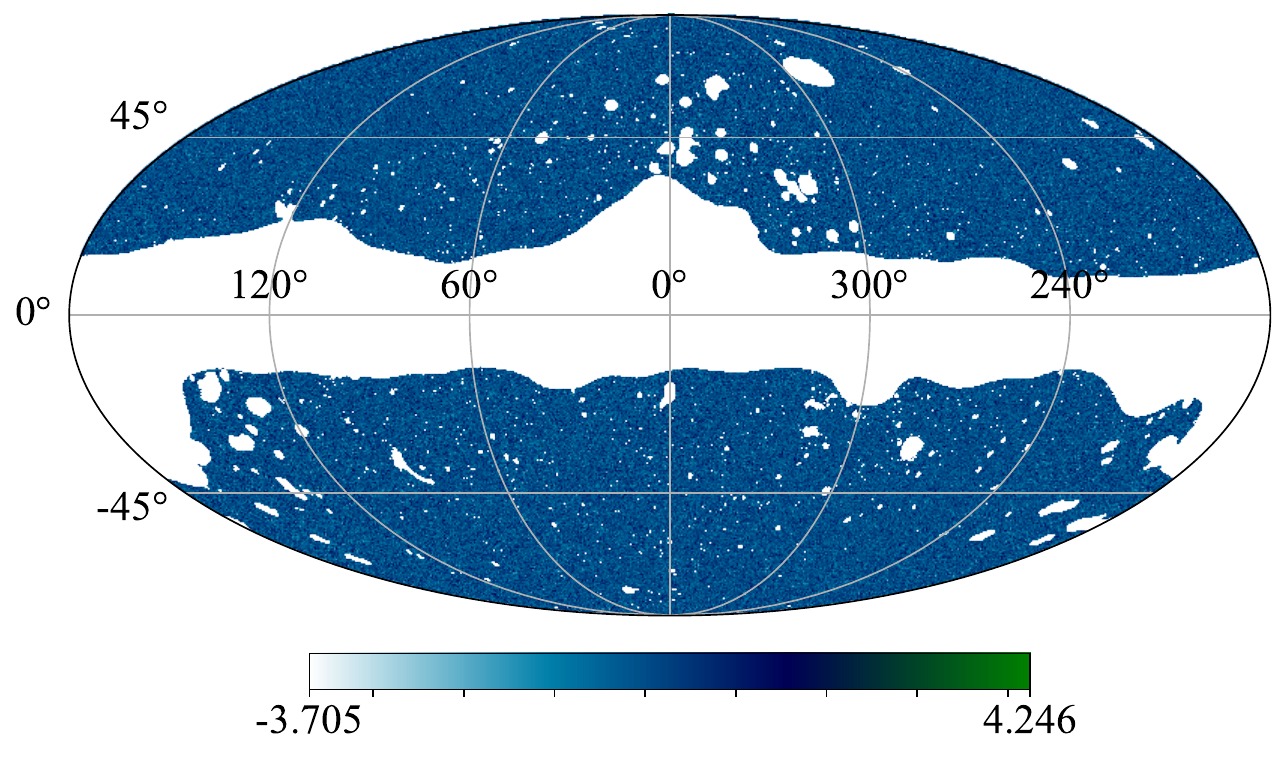}
    \caption{The {\it Planck} DR4 CMB lensing map is presented in Galactic coordinates using the \texttt{HEALPix} framework \citep{2005_Gorski_healpix}. This map has a resolution parameter of $N_{\rm side}=512$.}
    \label{fig:CMB lensing}
\end{figure}
\subsection{CMB Weak Lensing}
We use an optimized lensing reconstruction of the latest {\it Planck} sky map \citep{2022_Carron_CMB_lensing}, processed through the NIPIPE pipeline \citep{2020_Planck_LFI_and_HFI_data_processing}, for cross-correlation with quasars. This approach yields up to a 20\% improvement in signal-to-noise ratio, thanks to several key enhancements: joint Wiener filtering of the CMB temperature and polarization maps to improve the E-mode recovery; Wiener filtering of the lensing convergence map reconstructed from quadratic estimators to provide optimal sky weighting \citep{2019_Mirmelstein_optimal_filtering}; inhomogeneous filtering for the GMV estimator; and the use of 600 Monte Carlo simulations to reduce the statistical uncertainties. In addition, a novel technique is employed to construct a realization-dependent covariance matrix that accounts for the spatially varying noise across the sky. The full methodology is described in \cite{2022_Carron_CMB_lensing}.

While the ground-based experiments are rapidly improving the CMB lensing reconstructions on small angular scales such as the Atacama Cosmology Telescope (ACT; \cite{2017_sherwin_Atacama_lensing,2020_Darwish_ACT,2024_Qu_ACTDR6,2025_Farren_ACT_unWISE}) and the South Pole Telescope (SPT; \cite{2019_Wu_SPTpol,2020_Bianchini_SPTPOL_lensing,2023_Pan_SPT3G2018,2025_Ge_SPT3G2019_2020}), the {\it Planck} CMB lensing remains the only full-sky dataset in the foreseeable future. It uniquely enables the reconstruction of the largest-scale lensing modes \citep{2023_Alonso_quaia_structure_growth}, making it especially well-suited for cross-correlation studies with wide-area, high-redshift quasar samples, thereby enhancing the cosmological constraints.

Figure \ref{fig:CMB lensing} shows the CMB lensing convergence map $\kappa(\hat{\mathbf{n}})$ and associated mask, both obtained from the public {\it Planck} legacy archive. Following \cite{2023_Alonso_quaia_structure_growth}, we rotate the spherical harmonic coefficients of the lensing map into equatorial coordinates, truncate them to $\ell_{\mathrm{max}} = 3N_{\mathrm{side}} - 1$, and perform an inverse spherical harmonic transform to produce a \texttt{HEALPix} map with $N_{\mathrm{side}} = 512$. The lensing mask is  rotated and apodized using a C1 kernel to suppress the edge effects.

\section{Methodology} \label{sec:meth}

\subsection{Selection Function}
Astronomical observations are invariably affected by the selection effects, whereby some objects in the true underlying population are missed during data collection and processing. These effects pose a significant challenge for the statistical studies aiming to obtain the intrinsic properties of astronomical populations. To address this, it is essential to characterize and correct for these biases using a well-defined selection function. A comprehensive introduction to constructing the selection functions is provided by \cite{2021_Rix_selection_function_tutorial}. Here, we offer a brief overview. 

In general, the selection function acts as a multiplicative factor linking the theoretical distribution of physical quantities to the distribution observed in real data:
\begin{equation}
\frac{d\Lambda(\boldsymbol{q})}{d\boldsymbol{q}}
=\mathcal{M}\left(\boldsymbol{q} \mid \boldsymbol{\Theta}\right) S(\boldsymbol{q}),
\end{equation}

where $d\Lambda(\boldsymbol{q})/d\boldsymbol{q}$ denotes the true distribution of the quantities of interest, expressed as a function of the specific properties $\boldsymbol{q}$. The term $\mathcal{M}(\boldsymbol{q} \mid \boldsymbol{\Theta})$ represents the model-predicted distribution, given parameters $\boldsymbol{\Theta}$. The selection function $S(\boldsymbol{q})$ quantifies the probability that an object with properties $\boldsymbol{q}$ is included in the observed sample. It accounts for observational incompleteness and is defined as a dimensionless function ranging between 0 and 1. In the ideal case of no selection bias (i.e., when $S(\boldsymbol{q}) = 1$), the observed distribution coincides with the theoretical model.

\begin{figure}
    \centering    \includegraphics[width=1.0\linewidth]{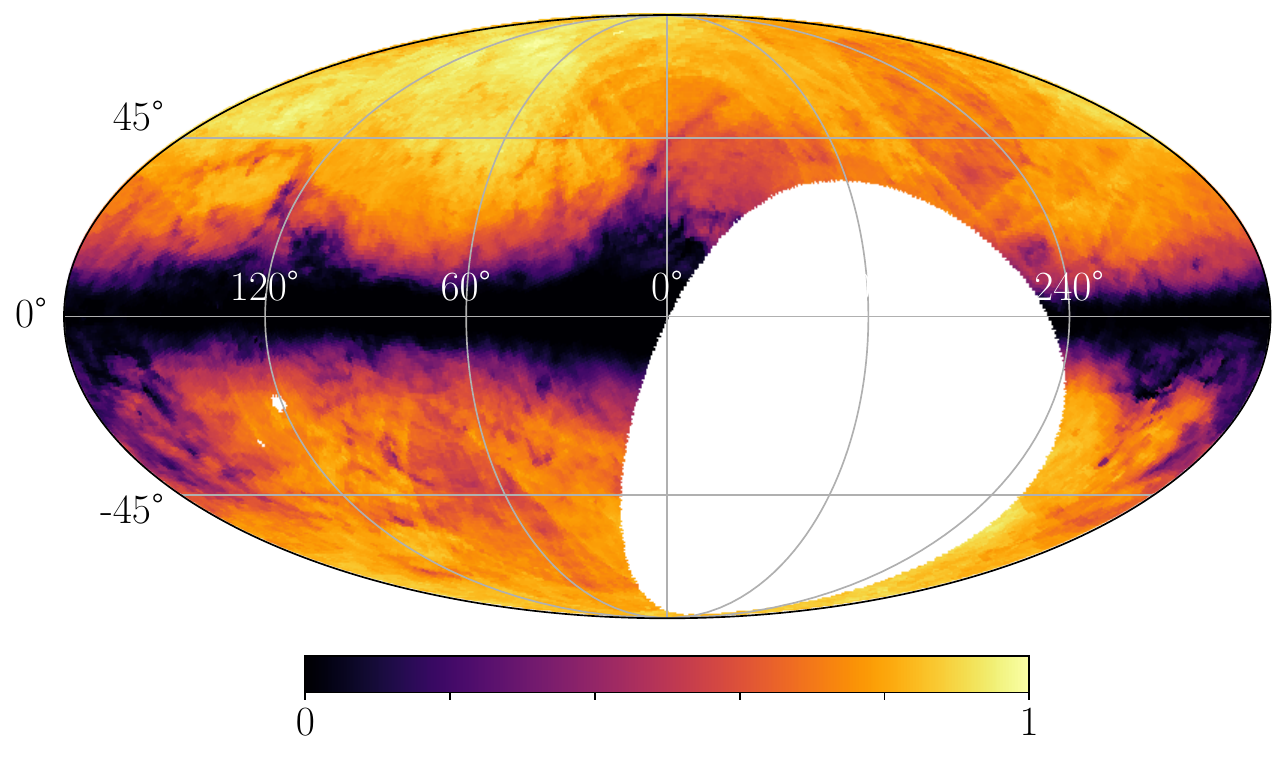}
    \caption{The selection function for the CatNorth catalog, without a magnitude cut, is presented in Galactic coordinates using the \texttt{HEALPix} framework with a resolution of \( N_{\rm side} = 64 \). Here, \( S(\boldsymbol{\hat{n}}) = 0 \) indicates the absence of data, while \( S(\boldsymbol{\hat{n}}) = 1 \) signifies complete data coverage. Significant incompleteness is evident in the Galactic plane and certain high Galactic latitude regions.}
    \label{fig:selection function}
\end{figure}

As mentioned in the introduction, quasar candidate selections are affected by systematic effects, including stellar contamination and spatial variations in the survey depth due to the differences in detection efficiency. These selection effects can significantly impact the cosmological constraints on large scales. In this work, we focus on how the selection function varies with sky position, denoted by a unit vector $\mathbf{\hat{n}}$, so that $S(\boldsymbol{q})$ reduces to $S(\mathbf{\hat{n}})$. As shown in Figure~\ref{fig:catnorth}, the spatial distribution of the CatNorth catalog is strongly modulated by the dust extinction, especially near the Galactic plane. Additional systematics, such as Gaia's scanning pattern, are also visible at high Galactic latitudes. To mitigate these effects, we developed a method that incorporates 10 systematics templates and employs neural network to model the spatial selection function for the CatNorth catalog. This method effectively suppresses most large-scale artifacts induced by the systematics while preserving the fluctuations associated with the underlying large-scale structure. Further details are provided in Appendix~\ref{appendix: selection function}. The selection function is built with a resolution of $N_{\rm side}=64$.

The quasar overdensity field $\delta(\mathbf{\hat{n}})$, with a resolution of $N_{\rm side} = 512$, is derived from the observed quasar number density field $n_{\text{obs}}(\mathbf{\hat{n}})$ using the selection function $S(\mathbf{\hat{n}})$, upgraded from $N_{\rm side} = 64$ to $N_{\rm side} = 512$:
\begin{equation}
\delta(\mathbf{\hat{n}}) = \frac{n_{\text{obs}}(\mathbf{\hat{n}})}{S(\mathbf{\hat{n}})n_u(\mathbf{\hat{n}})}-1,
\label{eq:delta}
\end{equation}
where $n_u(\mathbf{\hat{n}})$ is a uniform number density field representing the average number density. Figure~\ref{fig:selection function} shows the selection function for the full CatNorth sample, representing the spatial completeness across the sky. Note that for any subcatalog of CatNorth, the neural network used to estimate the selection function must be retrained. 
Note that Equation \eqref{eq:delta} enhances weights for low-completeness pixels. Following the standard pixelized approach described in the \texttt{pymaster} \citep{2019_Alonso_NaMaster} documentation, the selection function serves as an reasonable mask to mitigate this effect. Furthermore, following \cite{2023_Alonso_quaia_structure_growth}, we exclude pixels with $S(\mathbf{\hat{n}})< 0.5$ as they exhibit numerical instability when the selection function becomes too small. This approach is equivalent to applying a mask defined by $S(\mathbf{\hat{n}})> 0.5$.

In general, depth variations may lead to a selection function that is not purely angular. In our analysis, however, several factors mitigate the impact of such effects. Firstly, we impose a relatively bright magnitude cut when constructing our sample. This ensures that the catalog is close to complete over the vast majority of the survey footprint, thereby significantly reducing the sensitivity of our measurement to local depth fluctuations. Secondly, the sky regions close to the Galactic plane (where incompleteness and depth variations are strongest due to dust extinction and stellar contamination) are explicitly masked and excluded from the clustering analysis. Thirdly, there currently exists no substantially deeper and more complete ground truth catalog overlapping our footprint from which a reliable depth-dependent selection function could be empirically constructed. Therefore, in practice, a robust  depth-dependent correction cannot be derived without introducing additional and poorly constrained modeling assumptions. Given these considerations, we adopt the standard approach of modeling the selection function as purely angular, which should be a good approximation for our bright and nearly complete sample after masking.

\subsection{Angular Power Spectrum Analysis}\label{subsec: Cell}

We compute the theoretical angular power spectra for both auto-correlation and cross-correlation using the Core Cosmology Library (CCL) Python package \texttt{pyccl}. The theoretical framework implemented in \texttt{pyccl} is based on the methodology described in \citet{Chisari_2019_PYCCL}, and we provide a brief overview of the key aspects relevant to our analysis.

For two tracers $a$ and $b$, their angular cross-power spectrum $C^{ab}_\ell$ is defined in terms of  the ensemble average of spherical harmonic coefficients:
\begin{equation}
\left\langle a_{\ell m} b_{\ell' m'}^* \right\rangle = C^{ab}_\ell \, \delta_{\ell \ell^{\prime}} \delta_{m m^{\prime}},
\label{eq:cl_definition}
\end{equation}
where $\delta_{ij}$ is the Kronecker delta. 

The direct computation of $C^{ab}_\ell$ involves a computationally intensive triple integral over radial comoving distances. This calculation can be significantly accelerated under the Limber approximation \citep{1954_Limber_approximation,2004_Adshordi_Limber_approximation}, which applies when both tracers have broad radial kernels and at high multipoles ($\ell \gtrsim 20$). The approximation replaces the spherical Bessel function $j_\ell(x)$ with a Dirac delta function:
\begin{equation}
j_{\ell}(x) \approx \sqrt{\frac{\pi}{2 \ell+1}} \, \delta\left(\ell+\frac{1}{2}-x\right).
\label{eq:limber}
\end{equation}

By defining the radial distance as \( \chi_\ell \equiv (\ell+1/2)/k \) for each \( k \) and \( \ell \), which corresponds to a redshift \( z_\ell \), the power spectrum can be expressed as  
\begin{equation}
C_{\ell}^{a b} = \frac{2}{2\ell+1} \int_0^{\infty} d k \, \mathcal{P}(k, z_\ell) \Delta_{\ell}^a(k) \Delta_{\ell}^b(k),
\end{equation}
where \( \mathcal{P}(k, z_\ell) \) is the dimensionless power spectrum of the primordial curvature perturbations, and \( \Delta_{\ell}^a(k) \) and \( \Delta_{\ell}^b(k) \) are the transfer functions.

\begin{figure}
    \centering    \includegraphics[width=1.0\linewidth]{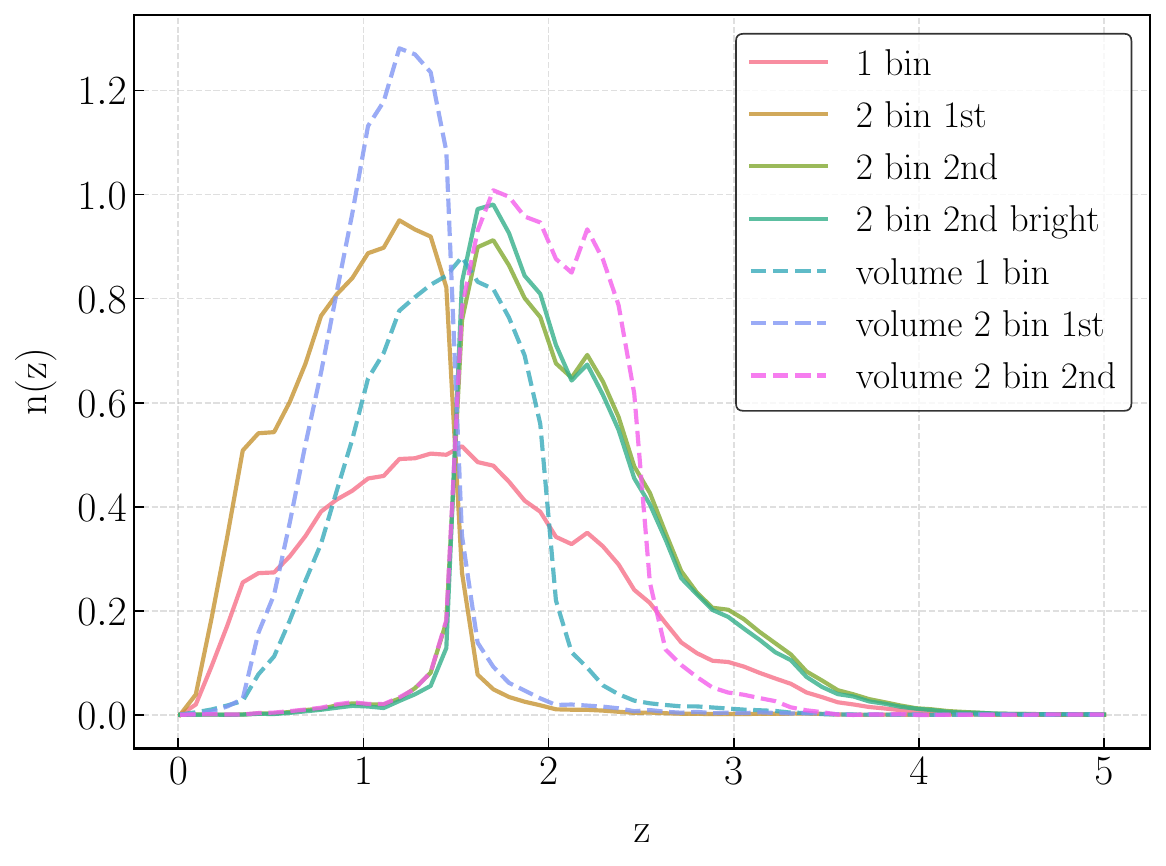}
    \caption{The normalized redshift distribution calibrated using the DIR method, for different subcatalogs in CatNorth.}
    \label{fig:nz}
\end{figure}

For the quasar tracer in the form of an overdensity field \( \delta(\mathbf{\hat{n}}) \), the transfer function is given by:  
\begin{equation}
    \Delta^{\delta}_\ell(k) = \Delta^{\mathrm{D}}_\ell(k) + \Delta^{\mathrm{RSD}}_\ell(k) + \Delta^{\mathrm{M}}_\ell(k),
    \label{Delta_ell_delta}
\end{equation}
where \( \Delta^{\mathrm{D}}_\ell(k) \) is the standard density term, which is proportional to the matter density, and can be expressed as:
\begin{equation}  
\Delta_{\ell}^{\mathrm{D}}(k)=p\left(z_{\ell}\right) b\left(z_{\ell}\right) H\left(z_{\ell}\right),
\label{Delta_ell_D}
\end{equation} 
where $H(z_\ell)$ is the Hubble parameter representing the expansion rate of the universe at redshift $z_\ell$ and $ p(z_\ell) $ represents the redshift distribution of the quasar sample. In this work, we use $\texttt{z}_{\texttt{ph}}$ from CatNorth and divide the catalog into two redshift bins based on $\texttt{z}_{\texttt{ph}}$. Figure \ref{fig:nz} shows the redshift distributions for different binning schemes, all calibrated using the direct calibration (DIR) method \citep{DIR_redshift_distribution}, which is implemented by cross-matching CatNorth with the SDSS DR16Q catalog \citep{2020_Lyke_sdssdrq16} in a 10-dimensional magnitude space (Gaia \texttt{G}, \texttt{BP}, \texttt{RP}, Pan-STARRS1 \texttt{g}, \texttt{r}, \texttt{i}, \texttt{z}, \texttt{y}, and CatWISE2020 \texttt{W1}, \texttt{W2}). The effective redshift, $z_{\rm eff}$, associated with a given redshift bin is defined as
\begin{equation}
    z_{\rm eff} \equiv \frac{\int  zp(z)dz}{\int p(z) dz}.
    \label{eq:effective redshift}
\end{equation}
$ b(z_\ell) $ denotes the linear bias for quasars. Following \cite{2021_Garcia-Garcia_quasar_growth_history}, we adopt the following bias model:
\begin{equation}
    b(z) = b_g(0.278((1+z)^2-6.565)+2.393),
    \label{bias}
\end{equation}
where $b_g$ is a free parameter.
 We also test alternative bias models in Section~\ref{sec:results}.

\begin{figure}
    \centering    \includegraphics[width=1.0\linewidth]{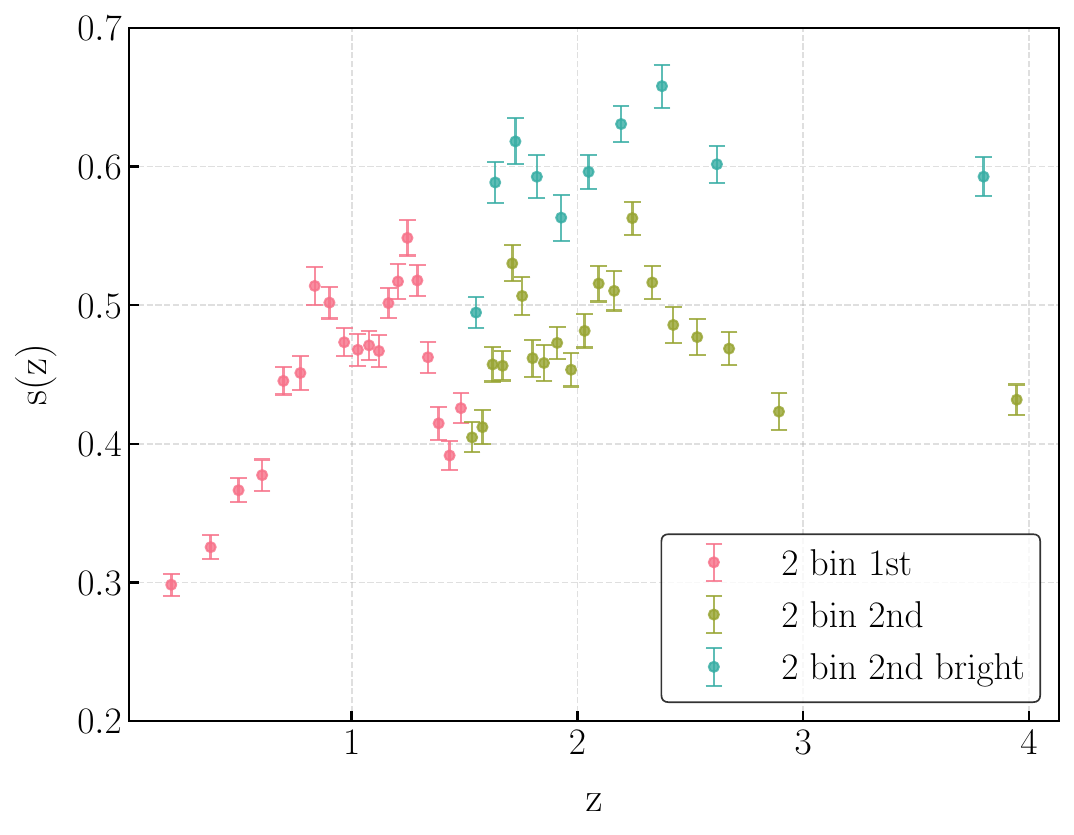}
    \caption{The logarithmic derivative of source counts with respect to the magnitude limit for different subcatalogs in CatNorth.}
    \label{fig:sz}
\end{figure}

\( \Delta^{\mathrm{RSD}}_\ell(k) \) in equation \eqref{Delta_ell_delta} represents the transfer function for the linear contribution from redshift-space distortions (RSD):
\begin{equation}
\begin{aligned}
& \Delta_{\ell}^{\mathrm{RSD}}(k)=\frac{1+8 \ell}{(2 \ell+1)^2} p\left(z_{\ell}\right) f\left(z_{\ell}\right) H\left(z_{\ell}\right) \\
& \quad\quad-\frac{4}{2 \ell+3} \sqrt{\frac{2 \ell+1}{2 \ell+3}} p\left(z_{\ell+1}\right) f\left(z_{\ell+1}\right) H\left(z_{\ell+1}\right),
\end{aligned}
\end{equation}
where \( f(z) \equiv d\ln[D(z)]/d\ln[a(z)] \) is the logarithmic growth rate, defined in terms of the linear growth factor of matter perturbations \( D(z) \) and the scale factor \( a(z) \). Note that \texttt{pyccl} employs a linear-theory relation between the matter overdensity and peculiar velocity fields, mediated by the scale-independent growth rate \( f(z) \). This approach may introduce inaccuracies when using narrow redshift bins. However, our analysis shows that the RSD effect has a negligible impact on the angular power spectrum predictions, implying that including RSD corrections does not significantly influence the cosmological parameter constraints.

\( \Delta^{\mathrm{M}}_\ell(k) \) in equation \eqref{Delta_ell_delta} represents the contribution from lensing magnification:
\begin{equation}
\begin{aligned}
&\Delta_{\ell}^{\mathrm{M}}(k) = 3 \Omega_{m} H_0^2 \frac{\ell(\ell+1)}{k^2} \frac{(1+z_{\ell})}{\chi_{\ell}} \\
&\quad \times \int_{z_\ell}^{\infty} dz^{\prime} \, p(z^{\prime}) \frac{2-5 s(z^{\prime})}{2} \frac{r(\chi^{\prime}-\chi_\ell)}{r(\chi^{\prime}) r(\chi_\ell)},
\end{aligned}
\label{eq:mag}
\end{equation}
where \( r(\chi) \) is the angular comoving distance, and \( s(z) \) is the logarithmic derivative of the number of sources with respect to a magnitude limit, defined as \( s(z) \equiv d\log_{10}N(<m)/dm \). \( N \) represents the cumulative number of sources. Figure \ref{fig:sz} shows the estimated \( s(z) \) values for different subcatalogs in CatNorth. 

For the CMB lensing tracer, the transfer function is given by:
\begin{equation}  
\begin{aligned}  
&\Delta_{\ell}^\kappa(k) = -\frac{\ell(\ell+1)}{2} \\  
&\quad \times \int_0^{\chi_*} \frac{dz}{H(z)} \frac{r(\chi_*-\chi)}{r(\chi) r(\chi_*)} T_{\phi+\psi}(k, z),
\end{aligned}  
\end{equation}  
where \( \chi_* \) denotes the comoving distance to the last-scattering surface, and \( T_{\phi+\psi}(k, z) \) represents the transfer function for the Newtonian-gauge scalar metric perturbations.

We estimate the observational angular power spectrum using the Python package \texttt{pymaster}, which is a Python implementation of the \texttt{NaMaster} library \citep{2019_Alonso_NaMaster}.  

We assume that the noise in the \texttt{pymaster} analysis is dominated by Poisson shot noise. For a discrete quasar sample, the shot noise contributes a scale-independent term to the quasar auto-correlation angular power spectrum, which can be written as
$$
N^{gg}_\ell = \frac{\langle S(\hat{\mathbf{n}})\rangle}{\bar{N}/\Omega_{\rm pix}},
$$
where $\langle S(\hat{\mathbf{n}})\rangle$ is the sky-averaged selection function, $\Omega_{\rm pix}$ is the solid angle of a \texttt{HEALPix} pixel, and $\bar{N}$ is the mean number density of quasars, defined as
$$
\bar{N} = \frac{\sum_p N_p}{\sum_p S(\hat{\mathbf{n}})_p},
$$
with $N_p$ denoting the quasar counts in pixel $p$. This shot-noise contribution is subtracted from the measured power spectrum to obtain an unbiased estimate of the clustering signal. Cosmic variance is not treated as a noise bias term in the power spectrum. Instead, it is fully accounted for in the covariance matrix of the bandpowers, following the standard Gaussian covariance expression implemented in \texttt{pymaster}.

\begin{figure}
    \centering    \includegraphics[width=1.0\linewidth]{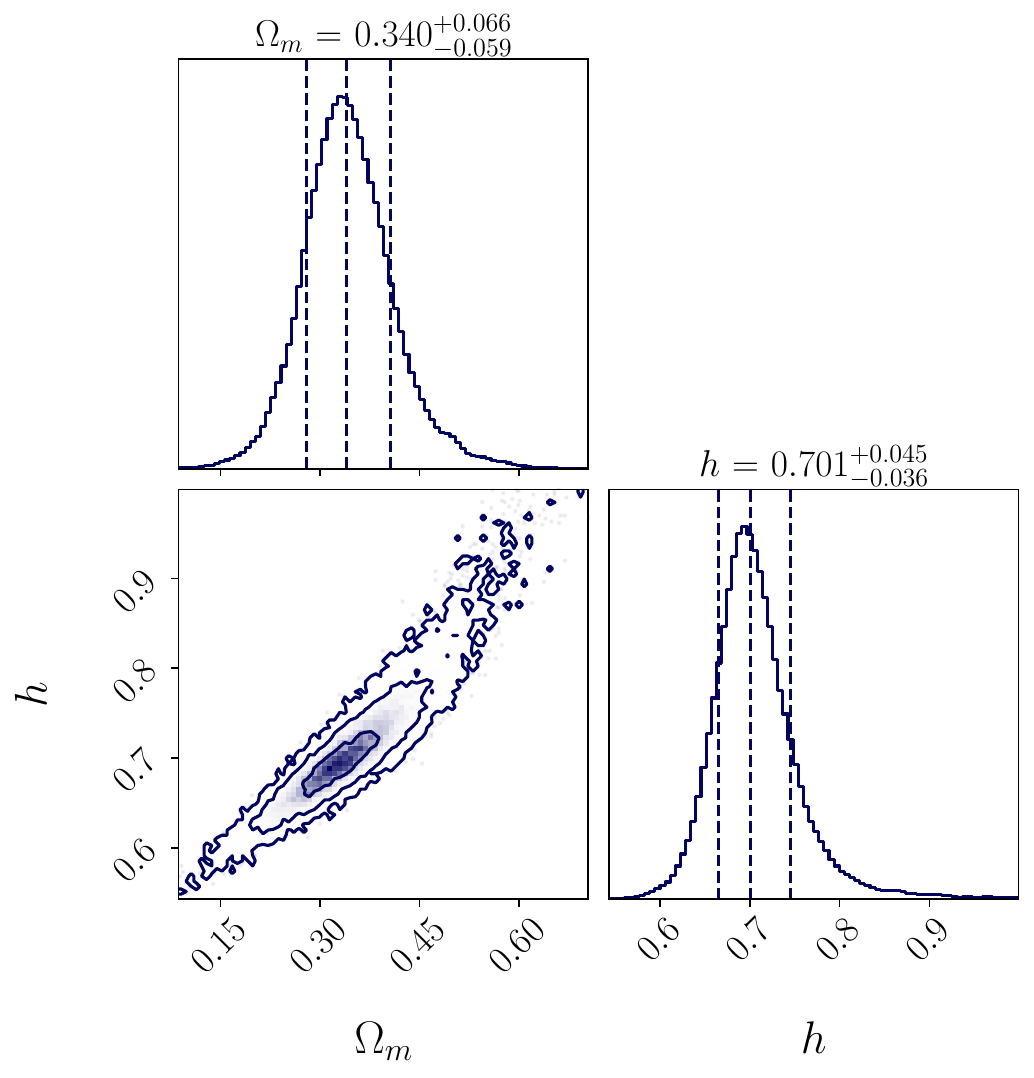}
    \caption{The BAO prior on $\Omega_m$ and $h$ used in this work.}
    \label{fig:bao}
\end{figure}
\subsection{Likelihood and Priors on Parameters}
We adopt a flat \(\Lambda\)CDM cosmology and treat the matter density parameter \(\Omega_m\), the reduced Hubble constant \(h\), and the amplitude of matter density fluctuations at \(8\, h^{-1} \mathrm{Mpc}\), \(\sigma_8\), as free parameters. Additionally, we include the quasar bias parameter \(b_g\), resulting in a total of four free parameters $\boldsymbol{\theta}\equiv\{\Omega_m, \sigma_8, h, b_g\}$. We fix the baryon density parameter \(\Omega_b = 0.0224/h^2\) and the spectral index of the primordial power spectrum \(n_s = 0.965\) based on {\it Planck} CMB anisotropy measurements \citep{2018_planck_cosmology_parameters}.

\begin{figure*}
    \centering
    \includegraphics[width=1.0\linewidth]{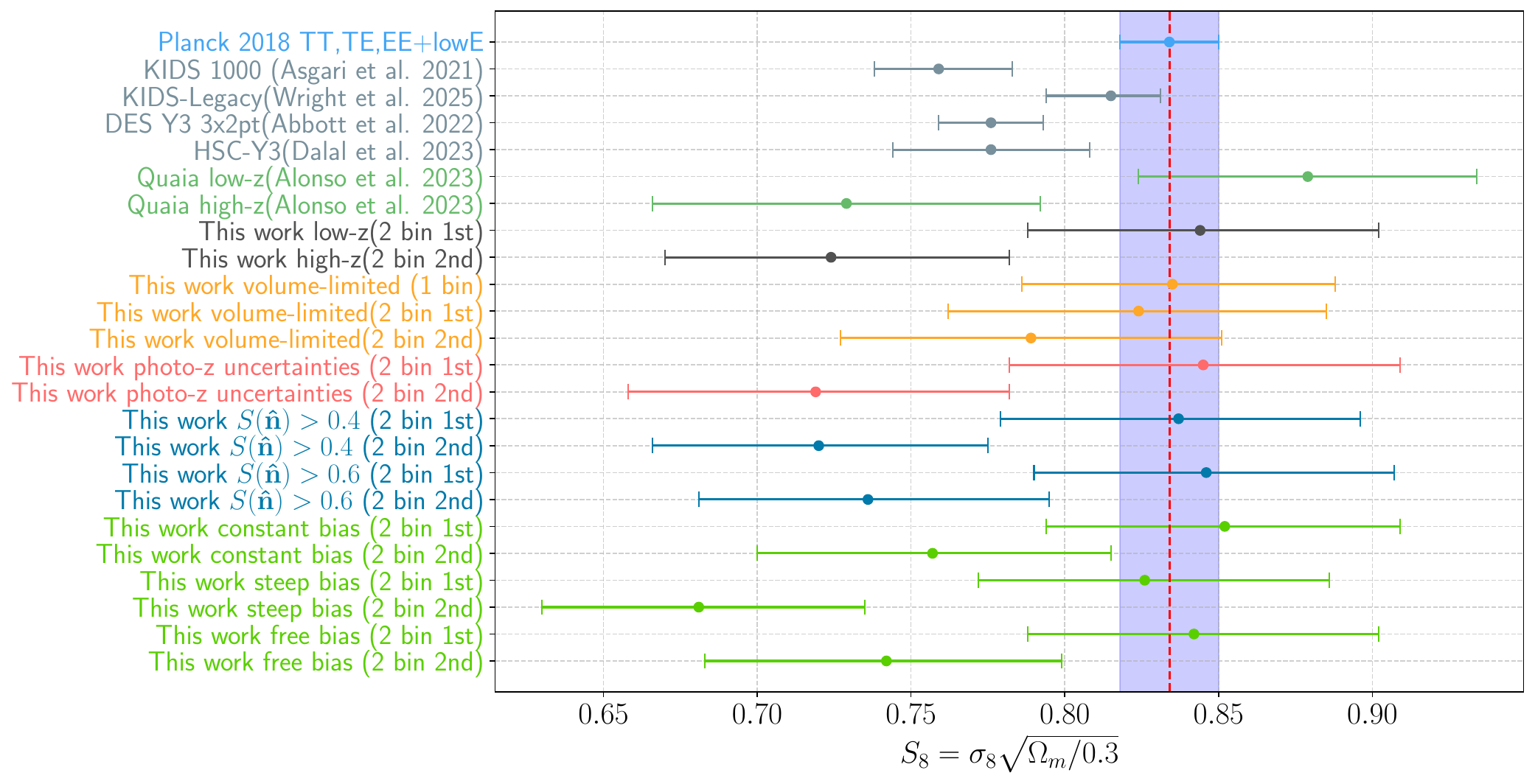}
    \caption{Constraints on the parameter $S_8$ from various cosmological probes. Points with horizontal error bars indicate the marginalized mean and $68\%$ confidence interval for each measurement. The results from this work are shown in the lower half of the plot, including constraints from the low- and high-redshift subsamples, three volume-limited subsamples, tests incorporating photometric redshift uncertainties, different masks, and different bias models. For comparison, we also show constraints from previous studies, including Planck 2018 CMB data, KiDS-1000 \citep{2021_kids1000_cosmic_shear}, DES Y3 3x2pt \citep{2022_DESY3_3_cross_2_cosmology}, HSC-Y3 \citep{2023_HSC_cosmology}, Quaia low- and high-redshift samples \citep{2023_Alonso_quaia_structure_growth}, and KiDS-Legacy \citep{2025_kids_legacy_cosmologicalconstraintscosmic}. The vertical red dashed line and shaded region represent the mean and $68\%$ confidence interval from the Planck 2018 TT,TE,EE+lowE analysis \citep{2018_planck_cosmology_parameters}.
    }
    \label{fig:treeplot}
\end{figure*}

According to Bayesian statistics, the log likelihood function can be expressed as:  
\begin{equation}  
\mathscr{L}^{ab}(\mathbf{C_\ell^{\mathrm{obs}}};\boldsymbol{\theta}) = -\frac{1}{2} \Delta^\mathrm{T} \Sigma^{-1} \Delta + \mathscr{C},
\end{equation}
where $\Delta = \mathbf{C_\ell^{\mathrm{obs}}} - \mathbf{C_\ell(\boldsymbol{\theta})}$ represents the residual between the observed and theoretical angular power spectra vector, and $\Sigma$ denotes the covariance matrix of the observed power spectrum, which is estimated as the Gaussian covariance implemented by \texttt{pymaster}. $\mathscr{C}$ is an arbitrary normalisation constant. To reduce the impact of cosmic variance on large angular scales, ensure the validity of the Limber approximation, and remain consistent with the linear bias approximation, we restrict the multipole range to $\ell_{\mathrm{min}} = 30$ and $\ell_{\mathrm{max}} = 0.15 \chi(z_{\rm eff}) - 0.5$, where $\chi(z_{\mathrm{eff}})$ denotes the comoving distance at the effective redshift $z_{\mathrm{eff}}$ for the specified subsample. The cosmological parameters adopted for the calculation of $\chi$ follow the \textit{Planck} 2018 values, which implies that the size of the data vector is fixed and determined solely by the effective redshift of the specific data sample. The unit of $\chi$ is $\mathrm{Mpc}$.


The priors for parameters \(\boldsymbol{\theta}\) are listed in Table \ref{tab:prior and best fit}. We apply a uniform prior for \(\Omega_m\) of \((0.05,0.7)\), \(\sigma_8\) of \((0.5,1.2)\), \(h\) of \((0.4,1.0)\), and \(b_g\) of \((0.1,3.0)\). To better break the degeneracy between \(\Omega_m\) and \(\sigma_8\), we follow \cite{2023_Alonso_quaia_structure_growth} and adopt a BAO prior for \(\Omega_m\) and \(h\) derived from the \(D_M\) (comoving diameter distance) and \(H\) (Hubble parameter) data of \cite{2017_Alam_BAO_prior} at \(z = 0.38, 0.51\), as well as the \(D_M\) and \(D_H\) (Hubble distance) data from \cite{2020_Gil_Marin_BAO_prior} at \(z = 0.698\) (see Figure \ref{fig:bao} for the prior contours).

Finally, we use the Python module \texttt{emcee}, an Affine Invariant Markov Chain Monte Carlo (MCMC) Ensemble Sampler \citep{emcee}, to survey the posterior distribution in the parameter space.

\begin{table*}
	\newcolumntype{l}{>{\raggedright}p{0.17\textwidth}}
	\newcolumntype{c}{>{\centering}p{0.09\textwidth}}
    \newcolumntype{d}{>{\centering}p{0.045\textwidth}}
	\newcolumntype{r}{>{\centering\arraybackslash}p{0.15\textwidth}}
    \centering
	   \caption{Best-Fitting Values of Cosmological Parameters and Physical Quantities.}
       \label{tab:prior and best fit} 
	   \renewcommand\arraystretch{1.2}
            \rowcolors{2}{gray!10}{white} 
            \begin{tabular}{lcccccddr}
			\toprule
                \toprule
                 Qty./Param.  & $\Omega_m$ & $\sigma_8$ & $h$  & $b_g$ & $S_8$ & $z_{\mathrm{eff}}$ & $f_{\rm sky}$ & $n^{\rm mask}_{\rm qso}/n^{\rm all}_{\rm qso}$ \\ \midrule
                Prior & $U(0.05, 0.7)$\\+BAO & $U(0.5,1.2)$ & $U(0.4,1.0)$+\\BAO  & $U(0.1,3.0)$ & ... & ... & ...& ...\\
                2 bin 1st &$0.344^{+0.025}_{-0.023}$&$0.789^{+0.056}_{-0.054}$&$0.707^{+0.040}_{-0.032}$&$1.240^{+0.122}_{-0.115}$&$0.844^{+0.058}_{-0.056}$&0.964& 0.559& 518037/574411\\
                2 bin 2nd&$0.337^{+0.022}_{-0.023}$&$0.684^{+0.050}_{-0.048}$&$0.697^{+0.034}_{-0.031}$&$1.305^{+0.119}_{-0.111}$&$0.724^{+0.058}_{-0.054}$ & 2.149& 0.551 & 515532/574410\\
                2 bin joint &$0.346^{+0.024}_{-0.021}$&$0.731^{+0.038}_{-0.037}$&$0.697^{+0.030}_{-0.030}$&$1.355^{+0.101}_{-0.092}$\newline $1.211^{+0.086}_{-0.081}$&$0.785^{+0.042}_{-0.041}$ & ...& 0.559 &1033569/1148821\\
                2 bin 1st(Quaia) &$0.336^{+0.020}_{-0.022}$&$0.831^{+0.046}_{-0.046}$&...&...&$0.879^{+0.055}_{-0.055}$ & 1.004&0.742 &618137/647751\\
                2 bin 2nd(Quaia) &$0.344^{+0.022}_{-0.024}$&$0.680^{+0.051}_{-0.051}$&...&...&$0.729^{+0.063}_{-0.063}$ & 2.079 & 0.676&601216/647751\\
                2 bin joint(Quaia) &$0.343^{+0.017}_{-0.019}$&$0.766^{+0.034}_{-0.034}$&...&...&$0.819^{+0.042}_{-0.042}$ & ...&0.742 &1219353/1295502\\

                2 bin 2nd bright&$0.330^{+0.025}_{-0.024}$&$0.763^{+0.073}_{-0.067}$&$0.692^{+0.037}_{-0.034}$&$1.097^{+0.168}_{-0.148}$&$0.802^{+0.080}_{-0.077}$&2.128& 0.556 &261363/287206\\
                volume 1 bin &$0.337^{+0.022}_{-0.022}$&$0.788^{+0.049}_{-0.047}$&$0.703^{+0.033}_{-0.031}$&$1.147^{+0.088}_{-0.084}$&$0.835^{+0.053}_{-0.049}$&1.415&0.569 &771827/863713\\
                volume 2 bin 1st & $0.334^{+0.023}_{-0.023}$&$0.781^{+0.059}_{-0.056}$&$0.688^{+0.033}_{-0.031}$&$1.199^{+0.115}_{-0.106}$&$0.824^{+0.061}_{-0.062}$&1.129& 0.579&479424/528982\\
                volume 2 bin 2nd & $0.327^{+0.023}_{-0.023}$&$0.756^{+0.056}_{-0.057}$&$0.683^{+0.032}_{-0.031}$&$1.109^{+0.101}_{-0.094}$&$0.789^{+0.062}_{-0.062}$&2.001&0.568 &469686/517042\\
                volume 2 bin joint & $0.330^{+0.022}_{-0.020}$&$0.763^{+0.042}_{-0.041}$&$0.678^{+0.031}_{-0.029}$&$1.224^{+0.087}_{-0.081}\newline 1.096^{+0.076}_{-0.069}$&$0.800^{+0.045}_{-0.045}$&...& 0.579&949110/1046024\\
                photo-z uncertainties 2 bin 1st& $0.341^{+0.025}_{-0.024}$&$0.793^{+0.060}_{-0.060}$&$0.704^{+0.038}_{-0.033}$&$1.221^{+0.152}_{-0.140}$&$0.845^{+0.064}_{-0.063}$&0.964& 0.559&518037/574411\\
               photo-z uncertainties 2 bin 2nd & $0.336^{+0.022}_{-0.023}$&$0.680^{+0.058}_{-0.054}$&$0.698^{+0.033}_{-0.031}$&$1.316^{+0.180}_{-0.152}$&$0.719^{+0.063}_{-0.061}$&2.149& 0.551 & 515532/574410\\
                $S(\mathbf{\hat n})>0.4$ 2 bin 1st & $0.345^{+0.026}_{-0.022}$&$0.780^{+0.056}_{-0.055}$&$0.710^{+0.041}_{-0.034}$&$1.274^{+0.127}_{-0.114}$&$0.837^{+0.059}_{-0.058}$&0.964& 0.585 & 531352/574411\\
                $S(\mathbf{\hat n})>0.4$ 2 bin 2nd & $0.337^{+0.022}_{-0.021}$&$0.679^{+0.051}_{-0.047}$&$0.700^{+0.033}_{-0.030}$&$1.326^{+0.117}_{-0.112}$&$0.720^{+0.055}_{-0.054}$&2.149& 0.577 & 528257/574410\\
                $S(\mathbf{\hat n})>0.6$ 2 bin 1st & $0.343^{+0.025}_{-0.024}$&$0.793^{+0.055}_{-0.055}$&$0.706^{+0.036}_{-0.034}$&$1.224^{+0.123}_{-0.112}$&$0.846^{+0.061}_{-0.056}$&0.964& 0.528 & 499937/574411\\
                $S(\mathbf{\hat n})>0.6$ 2 bin 2nd & $0.337^{+0.023}_{-0.023}$&$0.695^{+0.052}_{-0.049}$&$0.698^{+0.034}_{-0.031}$&$1.267^{+0.115}_{-0.109}$&$0.736^{+0.059}_{-0.055}$&2.149& 0.516 & 495287/574410\\    
                2 bin 1st constant bias & $0.351^{+0.028}_{-0.024}$&$0.785^{+0.059}_{-0.054}$&$0.718^{+0.044}_{-0.035}$&$1.998^{+0.214}_{-0.191}$&$0.852^{+0.057}_{-0.058}$&0.964& 0.559& 518037/574411\\
                2 bin 1st steep bias & $0.345^{+0.025}_{-0.023}$&$0.770^{+0.057}_{-0.052}$&$0.708^{+0.039}_{-0.034}$&$1.525^{+0.150}_{-0.145}$&$0.826^{+0.060}_{-0.054}$&0.964& 0.559& 518037/574411\\
                2 bin 1st free bias & $0.346^{+0.025}_{-0.023}$&$0.784^{+0.057}_{-0.053}$&$0.710^{+0.040}_{-0.033}$&$...$&$0.842^{+0.060}_{-0.054}$&0.964& 0.559& 518037/574411\\
                2 bin 2nd constant bias$^{*}$ & $0.337^{+0.024}_{-0.022}$&$0.712^{+0.053}_{-0.048}$&$0.703^{+0.035}_{-0.031}$&$3.822^{+0.342}_{-0.324}$&$0.757^{+0.058}_{-0.057}$&2.149& 0.551 & 515532/574410\\
                2 bin 2nd steep bias & $0.332^{+0.023}_{-0.023}$&$0.648^{+0.048}_{-0.046}$&$0.690^{+0.032}_{-0.031}$&$1.032^{+0.093}_{-0.086}$&$0.681^{+0.054}_{-0.051}$&2.149& 0.551 & 515532/574410\\
                2 bin 2nd free bias & $0.336^{+0.023}_{-0.023}$&$0.702^{+0.050}_{-0.051}$&$0.699^{+0.036}_{-0.031}$&...&$0.742^{+0.057}_{-0.059}$&2.149& 0.551 & 515532/574410\\
                                
			\bottomrule
	   \end{tabular}

        \begin{flushleft}
        \footnotesize
        \textbf{Notes:} $\Omega_m$ denotes the matter density parameter, $\sigma_8$ represents the amplitude of matter fluctuations, $h \equiv H_0/(100\ \mathrm{km\ s^{-1}\ Mpc^{-1}})$ is the reduced Hubble constant, and $S_8 \equiv \sigma_8\sqrt{\Omega_m/0.3}$ is the combined parameter. The effective redshift $z_{\mathrm{eff}}$ is defined in Eq.~\eqref{eq:effective redshift}, $f_{\mathrm{sky}}$ indicates the sky coverage fraction, while $n^{\mathrm{mask}}_{\mathrm{qso}}/n^{\mathrm{all}}_{\mathrm{qso}}$ denotes the fraction of quasar candidates within the mask region relative to the total number of quasar candidates in the subsample.\\
        *For the ``2 bin 2nd constant bias'' case, we adopt a uniform prior $U(0.1,5.0)$ on the bias amplitude $b_g$, motivated by its relatively large best-fit value.
        \end{flushleft}
\end{table*}

\begin{figure}
    \centering
    \includegraphics[width=1.0\linewidth]{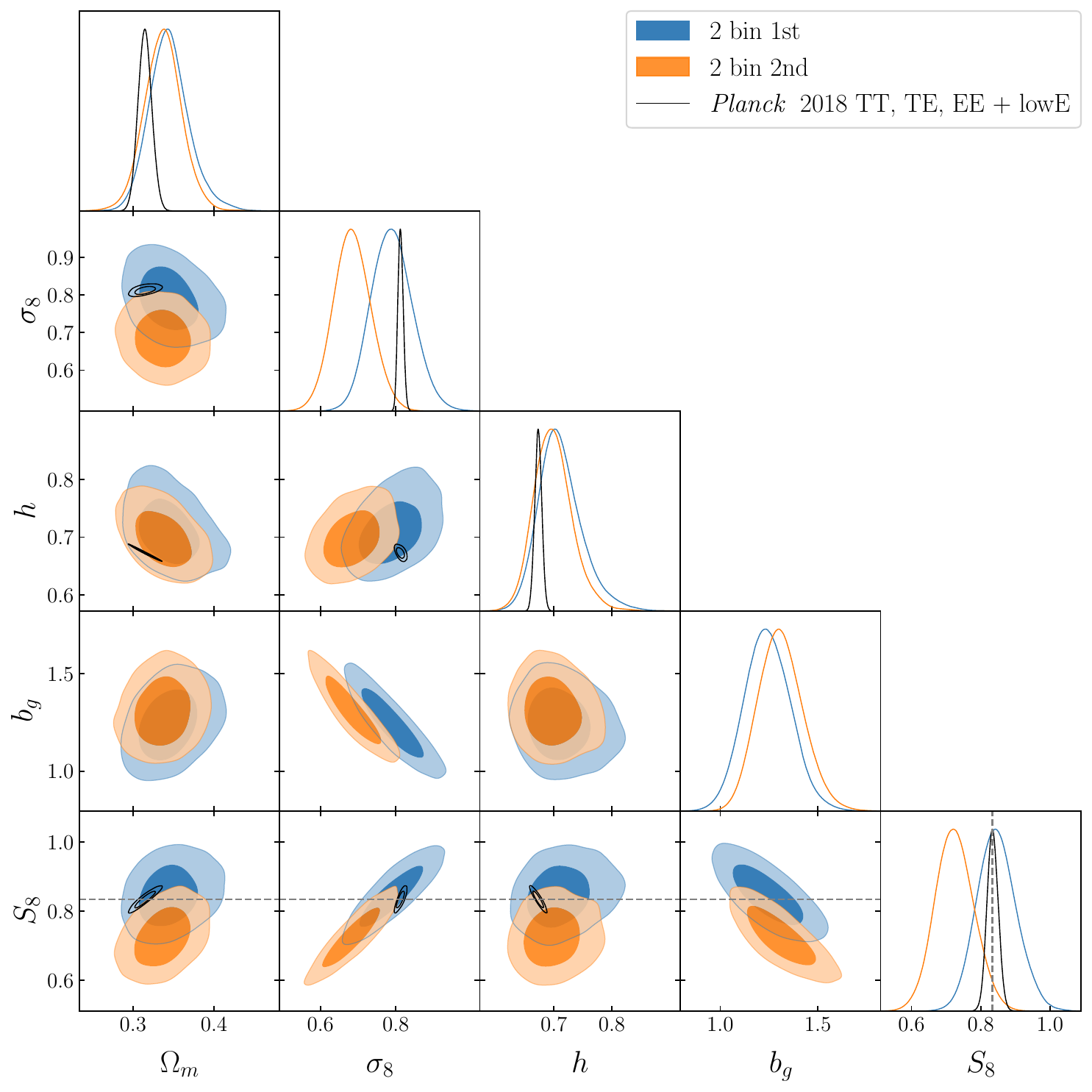}
    \caption{The 68$\%$ and 95$\%$ confidence contours for cosmological parameter constraints obtained in this work are shown under different binning schemes for the flux-limited samples. The smallest contour represents the constraints from {\it Planck} 2018 TT, TE, EE + lowE, with the best-fit value of $S_8=0.834\pm0.016$ indicated by the black solid line.
    }
    \label{fig:corner}
\end{figure}

\begin{figure}
    \centering
    \includegraphics[width=1.0\linewidth]{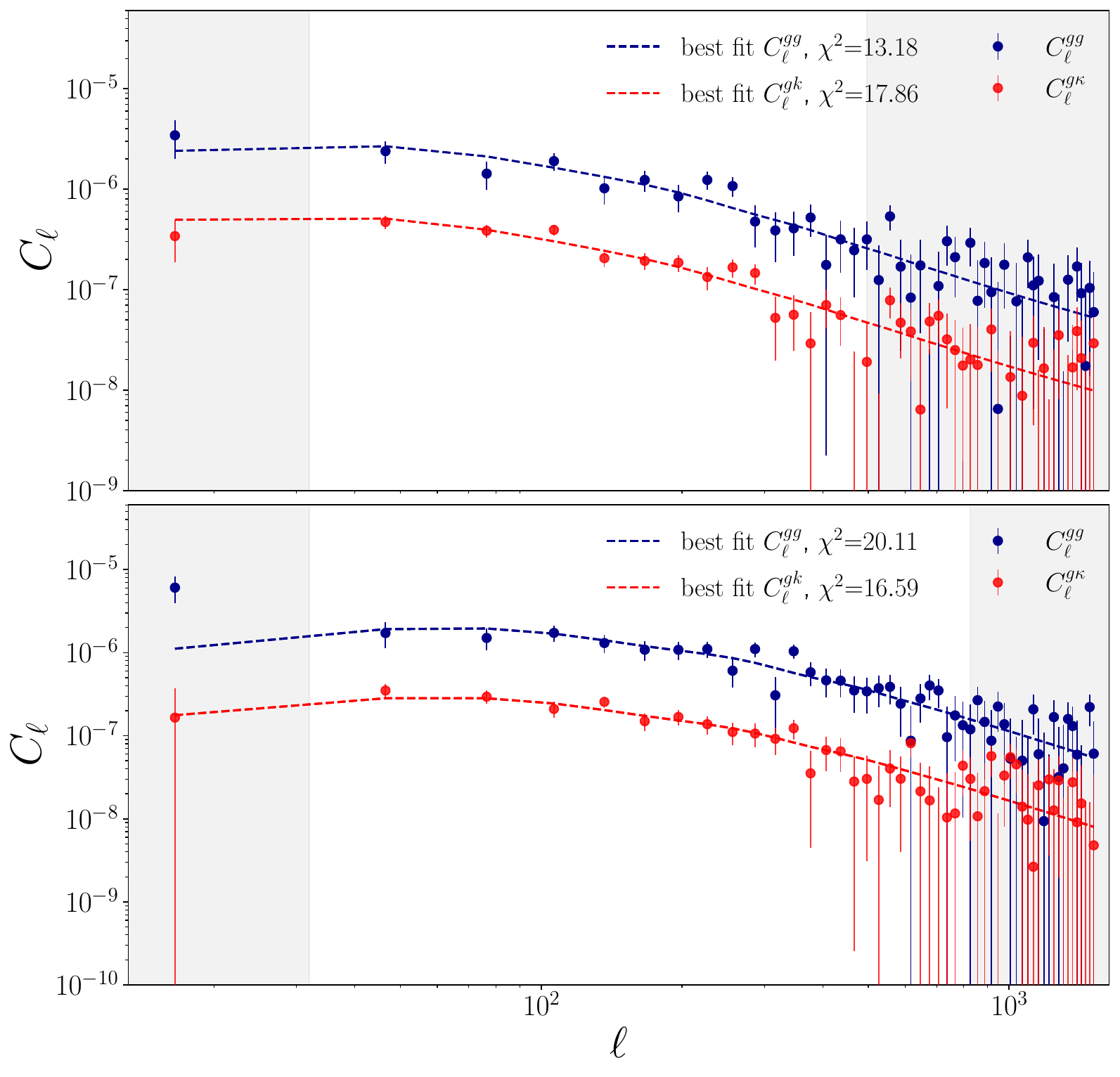}
    \caption{Two bin angular power spectrum measurements using \texttt{pymaster}. $C_\ell^{gg}$ represents the quasar auto-correlation power spectrum, while $C_\ell^{g\kappa}$ denotes the quasar-CMB lensing cross-correlation power spectrum. The top panel shows results for the low-redshift bin, and the bottom panel for the high-redshift bin. Data on the largest scales (left gray regions) were excluded due to cosmic variance and the breakdown of the Limber approximation, while nonlinear small-scale data (right grey regions) were omitted because: (1) pymaster cannot reliably estimate power spectra beyond $2*N_{\rm side}$ scales, and (2) nonlinear regime modeling remains challenging. }
    \label{fig:cells}
\end{figure}

\section{Results and Discussion} \label{sec:results}

In this section, we present the estimated parameters and discuss potential biases. To enhance clarity, each paragraph begins with a bolded statement summarizing its key focus.

\textbf{(1) Joint constraint results for the 2 bin cases.} For the convenient comparison with previous studies, we adopted a binning strategy similar to that of \cite{2023_Alonso_quaia_structure_growth}, who used the Quaia quasar candidate catalog \citep{2024_Storey-Fisher_quaia_catalog} to constrain structure growth. Specifically, we selected sources with $\texttt{phot\_g\_mean\_mag} < 20.5$ and divided them into two redshift bins based on $\texttt{z}_{\texttt{ph}}$, separated at $z = 1.5$, such that each bin contains an equal number of sources. Each bin was assigned an independent quasar bias parameter $b_g$. We then performed joint parameter constraints for the low- and high-redshift bins. The results, summarized in Table 1 under ``2 bin joint", yield $S_8 = 0.785^{+0.042}_{-0.041}$,  which is consistent with the previous large-scale structure measurements \citep{2021_kids1000_cosmic_shear,2021_kids1000_weak_lensing_galaxy_clustering,2023_HSC_cosmology,2022_DESY3_cosmic_shear_cosmology,2022_DESY3_3_cross_2_cosmology}. However, this estimate differs from that of \cite{2023_Alonso_quaia_structure_growth}, based on Quaia, which reported $S_8 = 0.819^{+0.042}_{-0.042}$
 , as well as from the {\it Planck} 2018 result of $0.834\pm0.016$ \citep{2018_planck_cosmology_parameters}.

\textbf{(2) Separate constraint results for the 2 bin cases.} We further independently performed parameter constraints for the low-redshift and high-redshift bins, rather than conducting a joint constraint. The results are shown in Table 1, labeled as ``2 bin 1st" and ``2 bin 2nd". For the low-redshift bin (``2 bin 1st"), we obtained $ S_8 = 0.844^{+0.058}_{-0.056} $, which is consistent with the {\it Planck} 2018 result of $ S_8 = 0.834\pm0.016$ \citep{2018_planck_cosmology_parameters}. In contrast, the high-redshift bin (``2 bin 2nd") yielded a smaller $ S_8 $ value of $ 0.724^{+0.058}_{-0.054} $, as shown in Table 1. When comparing our results with those of \cite{2023_Alonso_quaia_structure_growth}, labeled as ``2 bin 1st (Quaia)" and ``2 bin 2nd (Quaia)", we find that both our low- and high-redshift bins exhibit a general shift toward lower values compared to the corresponding Quaia results, which are $S_8 = 0.879^{+0.055}_{-0.055}$ for the low-redshift bin and $S_8 = 0.729^{+0.063}_{-0.063}$ for the high-redshift bin.

\begin{figure}
    \centering
    \includegraphics[width=1.0\linewidth]{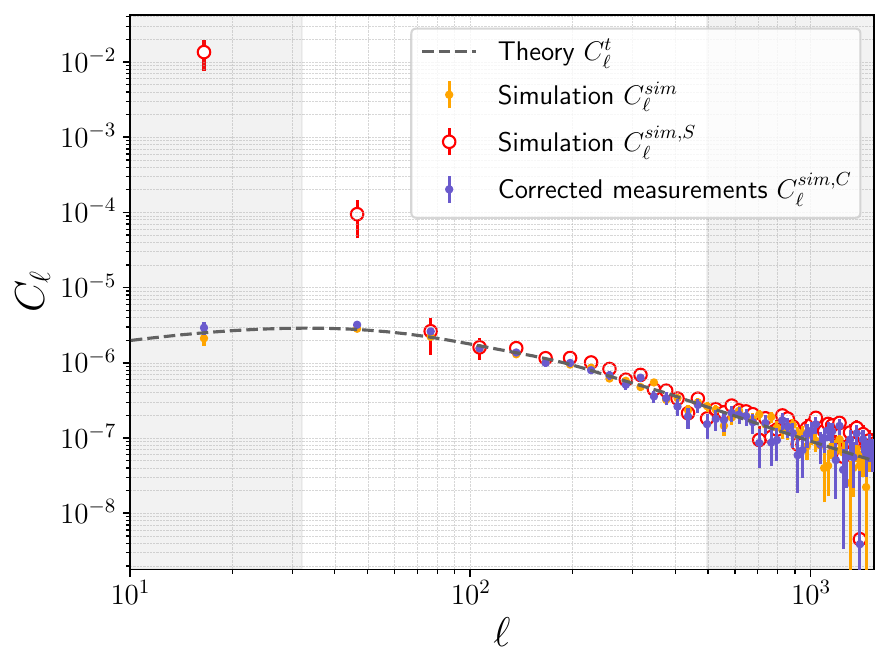}
    \caption{Testing oversuppression in a simulated dataset.
The dark gray dashed line represents the theoretical power spectrum used to generate the simulation. The orange data points with error bars show the power spectrum measured from simulations without selection effects. The red open circles represent the power spectrum measured from simulations with selection effects, which exhibit a noticeable large-scale excess. The blue data points show the corrected measurements, obtained by applying the selection function to mitigate observational systematics. The correction successfully brings the measurements back into agreement with the theoretical model, without introducing any obvious oversuppression.}
    \label{fig:oversuppresion}
\end{figure}

\begin{figure}
    \centering
    \includegraphics[width=1.0\linewidth]{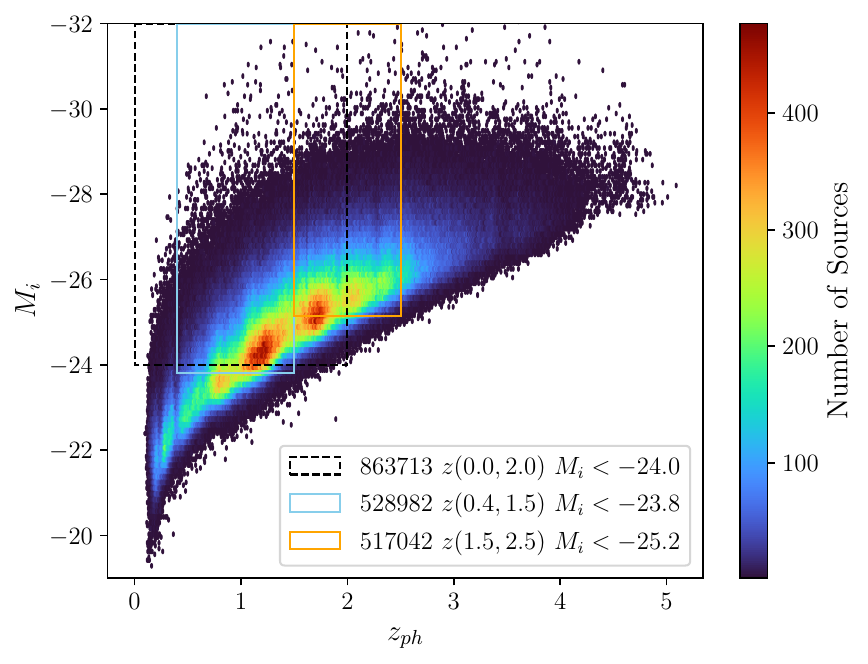}
    \caption{Binning strategy for volume-limited samples. The figure legends specify for each volume-limited sample: the number of objects, the redshift range, and the absolute magnitude range in the Pan-STARRS1 $\texttt{i}$-band, respectively.}
    \label{fig:vol_bin}
\end{figure}

\begin{figure*}
    \centering
    \includegraphics[width=1.0\linewidth]{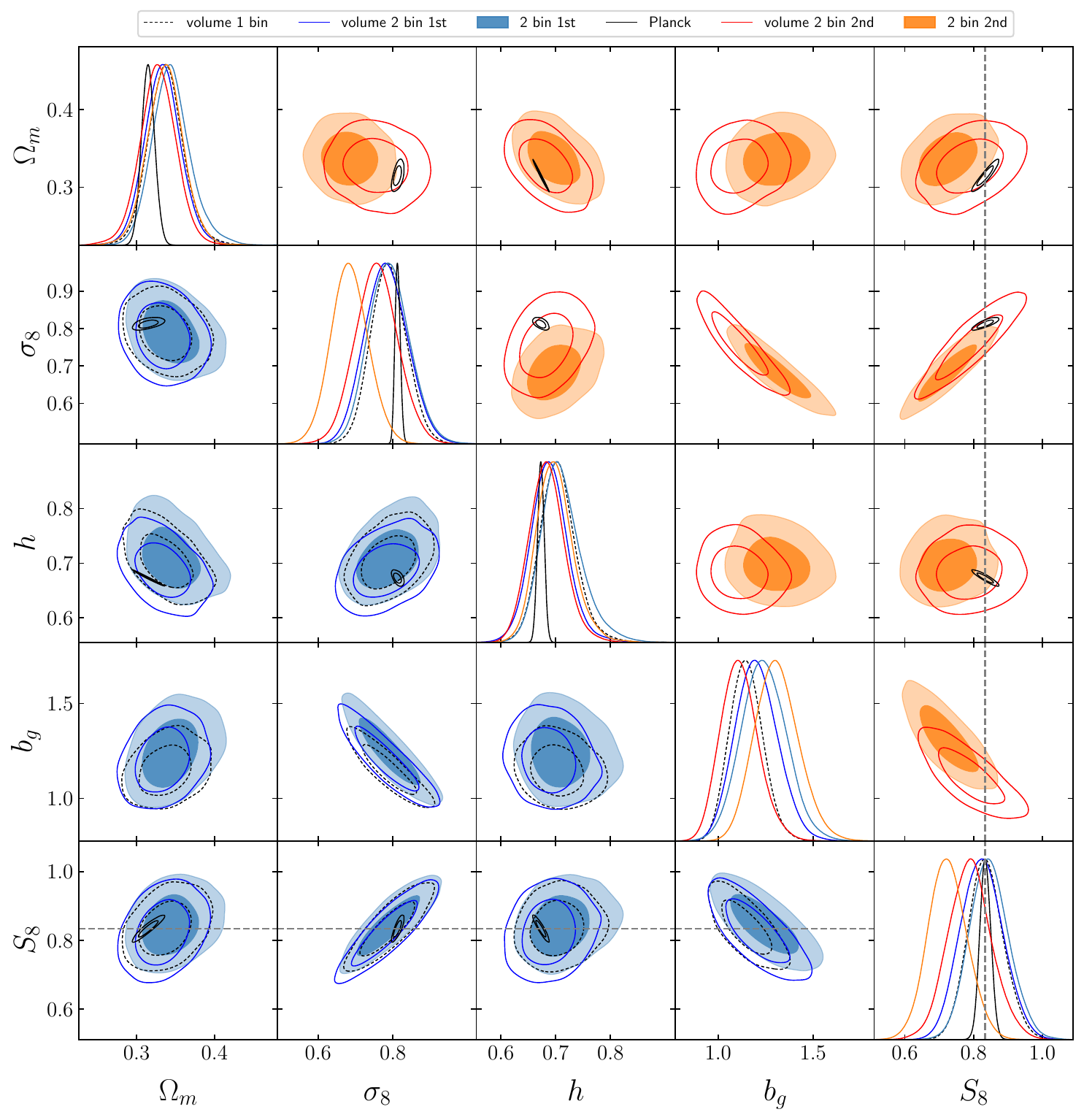}
    \caption{Similar to Figure \ref{fig:corner}, but obtained from the volume-limited samples. The bottom-left panels present the results for relatively low-redshift bins for both volume-limited and flux-limited samples, while the top-right panels show the results for relatively high-redshift bins. The smallest contour represents the constraints from the \textit{Planck} 2018 TT, TE, EE + lowE, with the best-fit value of $S_8=0.834\pm0.016$ indicated by the solid black line.
    }
    \label{fig:corner_vol}
\end{figure*}

\begin{figure}
    \centering
    \includegraphics[width=1.\linewidth]{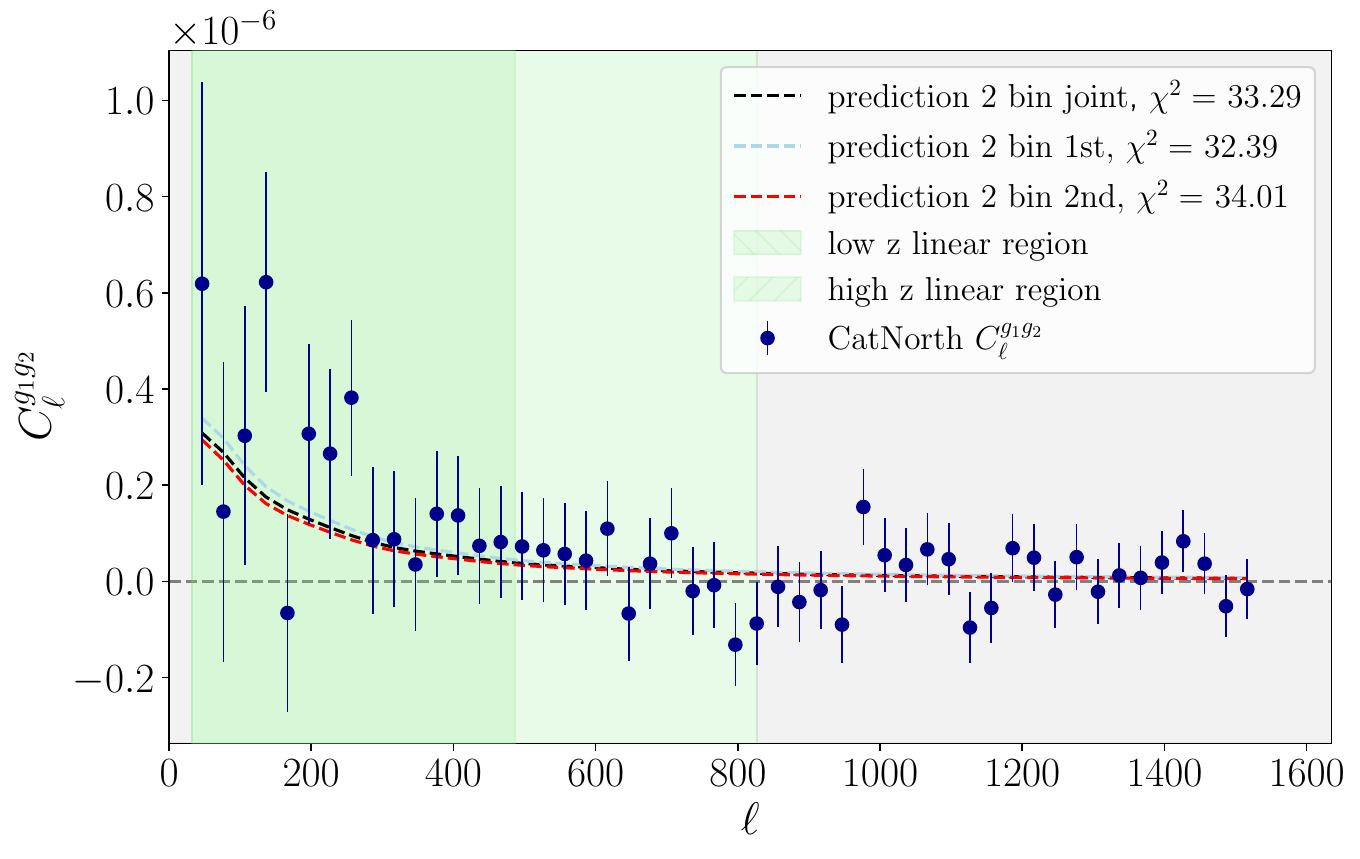}
    \caption{Cross angular power spectrum $C_\ell^{g_1g_2}$ between the low- and high-redshift quasar samples in the CatNorth catalog (blue points with error bars). The theoretical predictions are derived from the best-fit models to the auto-correlations of the individual redshift bins and their cross-correlations with the CMB lensing convergence map. The black dashed curve shows the joint prediction from the two-bin fit, while the light blue and red dashed curves correspond to predictions based solely on the low- and high-redshift bins, respectively. The shaded green regions indicate the linear scales adopted for the low-z and high-z samples. The consistency between the observed data points and theoretical predictions indicates that systematics mitigation is under control.}
    \label{fig:cross}
\end{figure}

\textbf{(4) Results from a bright high-redshift subsample of the 2 bin case.}
As an additional test, we further divided the high-redshift bin into two subsamples based on the apparent magnitude \texttt{phot\_g\_mean\_mag} with same number of objects. For the brighter subsample (labeled as ``2 bin 2nd bright" in Table \ref{tab:prior and best fit}), we repeated the parameter estimation and obtained a relatively large value of $S_8 = 0.802^{+0.080}_{-0.077}$, compared to the result from the full high-redshift bin (``2 bin 2nd").

\textbf{(5) Possible reasons for smaller $\boldsymbol{S_8}$ values compared to the previous Quaia-based work.} Thanks to the high-quality five-band photometry from Pan-STARRS1 and a machine learning-based photometric redshift estimation method, the CatNorth catalog is able to identify quasar candidates more reliably and reconstruct their redshift distributions more accurately than Quaia \citep{Fu_2024_CatNorth}. As discussed earlier, CatNorth also exhibits a higher source number density within the same sky coverage, indicating a higher level of completeness than Quaia. Although the inclusion of Pan-STARRS1 data introduces additional scan patterns and selection effects, our selection function effectively quantifies and corrects for these biases. 

In contrast, the Quaia selection function is constructed based on the unWISE source number density map and the exposure map, which are used as proxies for the WISE detection efficiency (see \cite{2024_Storey-Fisher_quaia_catalog}, Fig. 8). {However, the source number density does not always reliably reflect the true detection efficiency of the telescope. For example, in regions near the Galactic plane, the high source density is primarily due to the presence of bright stars, whereas the number of identified quasars in these areas remains relatively low. Therefore, we initially assume that the source number density is inversely proportional to the actual number of quasars. However, in regions near the ecliptic poles, which benefit from deeper exposures, the high source density may misleadingly imply ``fewer" quasars under this inverse-proportionality assumption—even though the increased exposure depth actually enables the detection of more quasars.} Relying solely on the source density maps can therefore lead to contradictory interpretations, and may result in an ineffective correction of selection effects—particularly in the ecliptic pole regions—introducing artifacts in the $C_\ell^{gg}$ and artificially overestimating the inferred $S_8$ values. By contrast, the use of the WISE median magnitude maps in CatNorth provides a more physically motivated tracer of observational systematics. These maps better capture the suppression of quasar detection in low Galactic latitude regions due to bright stars, as well as the enhanced quasar detection in deeply exposed regions near the ecliptic poles.

Another potential reason for the smaller value of $S_8$ compared to \cite{2023_Alonso_quaia_structure_growth} may lie in the relatively larger $s(z)$, as shown in Fig.~\ref{fig:sz}. Since CatNorth has a deeper survey depth than Quaia, when adopting the same magnitude cut of $G < 20.5$, CatNorth is closer to an ideal flux-limited sample. Owing to the intrinsic differences in the datasets, this results in a slightly larger measured $s(z)$. In contrast, \cite{2023_Alonso_quaia_structure_growth} reported $s(z < 1.47) = 0.388 \pm 0.004$ and $s(z > 1.47) = 0.420 \pm 0.005$ for the Quaia catalog.

In addition, the correction term $5s(z)-2$ in Eq.~\eqref{eq:mag} relies on the assumption of a strictly flux-limited sample. Deviations from this assumption can introduce biases in the measurement of $s(z)$. For instance, the process of photometric redshift estimation may induce selection effects that bias the magnification slope $s(z)$ \citep{2025_JianQin_magnification}. Unlike catalogs that require full-band detections to obtain photometric redshifts, CatNorth retains objects with missing photometric bands, thereby avoiding such selection effects.

Biases in the modeling of cosmic magnification can lead to systematic shifts in the theoretical angular power spectrum, with particularly strong impact on large scales and at high redshifts, and with greater influence on the cross-correlation $C_\ell^{g\kappa}$. This, in turn, may cause differences in the inferred value of $S_8$. Nevertheless, \cite{2023_Alonso_quaia_structure_growth} tested the extreme cases (e.g., adopting $s=0.2764$, which differs from their fiducial value by about $3\sigma$) and found that the resulting shift in $S_8$ was smaller than $1\sigma$, indicating that the impact of cosmic magnification on $S_8$ constraints remains relatively minor.

In summary, the higher data quality and completeness of the CatNorth catalog, combined with the more sophisticated treatment of selection effects in this work, are likely the key reasons why we obtain a lower $S_8$ value compared to Quaia.


\textbf{(6) Test for oversuppression.} A plausible concern is that the selection function might over-correct for systematics, suppressing power on large scales. To assess the robustness of our result and test for potential over-suppression, we perform a validation test using mock simulations.

We adopt the best-fit power spectrum from the ``2 bin first" configuration as the input theoretical power spectrum $C_\ell^t$ and generate the corresponding theoretical quasar overdensity map $\delta^t(\mathbf{\hat{n}})$. The mock simulations assume a mean number density of $\bar{n}^t=1$, producing a theoretical quasar number density map $n^t(\mathbf{\hat{n}})=\bar{n}^t(\delta^t(\mathbf{\hat{n}})+1)$. We then apply the selection function of the ``2 bin 1st" configuration to mimic observational selection effects, yielding the theoretical quasar number density map that includes systematics: $n^{t,S}(\mathbf{\hat{n}})=n^t(\mathbf{\hat{n}}) S(\mathbf{\hat{n}})$. 

Next, we treat the theoretical quasar number density in each pixel as the mean of a Poisson distribution and perform sampling. This generates two simulated maps: one without selection effects ($n^{\text{sim}}(\mathbf{\hat{n}})$, sampled from $n^{t}(\mathbf{\hat{n}})$) and one with selection effects ($n^{\text{sim},S}(\mathbf{\hat{n}})$, sampled from $n^{t,S}(\mathbf{\hat{n}})$). We then measure the angular power spectra of these two maps, obtaining $C_\ell^{\text{sim}}$ and $C_\ell^{\text{sim},S}$, respectively. 

Subsequently, we apply our correction procedure to compute the corrected angular power spectra $C_\ell^{\text{sim},\text{C}}$. The results are shown in Figure~\ref{fig:oversuppresion}. The selection effects introduce a prominent large-scale excess and a smaller but systematic excess on small scales. After applying our correction method, these excesses are effectively removed, with no significant over-suppression observed. We therefore conclude that our $S_8$ measurement is robust against potential biases introduced by the selection function correction.

It should be emphasized that the selection function approach, as a weighting method employed in this study, effectively corrects multiplicative errors but remains ineffective for additive errors. Regrettably, obtaining any reliable estimation of additive errors solely from the observational data proves fundamentally challenging, which may consequently lead to a potential overestimation of $C_\ell^{gg}$.

\textbf{(7) Possible reasons for smaller $\boldsymbol{S_8}$ in the high-redshift bin.}  
Both our results and those of \cite{2023_Alonso_quaia_structure_growth} consistently show that the $S_8$ value in the high-redshift bin is \textbf{smaller} than that in the low-redshift bin. This systematic discrepancy raises important questions about the reliability of high-redshift measurements. It appears unlikely that both CMB and low-redshift observations are simultaneously biased in such a way that only high-redshift measurements yield the correct $S_8$ value.

A more plausible explanation involves the inherent limitations in the completeness and quality of high-redshift quasar samples. Even with weighting methods such as selection functions, it remains challenging to fully recover unbiased cosmological information from these datasets. This is primarily due to the observational constraints: the large distances of high-redshift quasars make them susceptible to flux limits and photometric errors, which can introduce uncertainties in the color-based quasar selections and redshift estimations. In addition, the residual foreground contamination in CMB lensing maps may introduce systematic biases that suppress the inferred $S_8$ values at high redshift. The most likely source of such contamination is the Cosmic Infrared Background (CIB), since it affects CMB lensing more strongly at high redshift ($z\gtrsim 2$) and induces a negative bias in the power spectrum, thereby leading to a smaller inferred $S_8$ \citep{2023_Alonso_quaia_structure_growth}.

These factors suggest that the low-redshift results, which are based on more complete and robust data, may offer more reliable cosmological constraints and should be weighted more heavily in the related analyses.

\textbf{(8) Effects of a narrower binning strategy.}
While dividing the sample into narrower redshift bins could, in principle, provide better temporal resolution of cosmic evolution—commonly seen in many galaxy-based studies \citep{2021_kids1000_weak_lensing_galaxy_clustering,2021_kids1000_cosmic_shear,2023_Xu_S8}—this approach introduces significant systematic challenges in practice.

First, narrower bins reduce the number of objects per \texttt{HEALPix} pixel, degrading the selection function estimation accuracy. This may artificially bias the large-scale power spectrum and consequently yield unreliable $S_8$ measurements.

Second, narrower bins are more susceptible to photometric redshift uncertainties. The reconstructed redshift distribution becomes less robust and affects the theoretical auto-correlation power spectrum, while the cross-correlation remains relatively stable. To reconcile this discrepancy, the fitting algorithm may compensate by shifting the inferred $S_8$ value away from its true value.

In summary, narrow redshift bins may amplify systematic biases when working with limited data. We therefore adopt a two-bin configuration. For future quasar studies with limited data, we recommend using broader redshift bins to enhance the statistical reliability (by reducing Poisson noise) while decreasing the effects of redshift errors and selection effects.

\textbf{(9) Additional tests for the volume-limited samples.}
In principle, different binning strategies should yield consistent results when accounting for all selection effects. To validate the robustness of the $S_8$ constraints, we analyze several volume-limited samples. The binning strategy for the volume-limited samples is illustrated in Figure \ref{fig:vol_bin}. We first test a wide volume-limited sample with $z < 2$ and $M_i < -24$ (labeled as ``volume 1 bin" in Table \ref{tab:prior and best fit}), obtaining $S_8 = 0.835^{+0.053}_{-0.049}$, consistent with the {\it Planck} result within $1\,\sigma$. Furthermore, we test two volume-limited samples with narrower redshift binning (labeled as ``volume 2 bin 1st" and ``volume 2 bin 2nd") to test if $S_8$ varies with redshift, with the selection criteria $0.4 < z < 1.5,\quad M_i < -23.8$ and $1.5 < z < 2.5,\quad M_i < -25.2$, respectively. Similar to the previous 2 bin flux-limited sample results, the low-redshift sample yields $S_8 = 0.824^{+0.061}_{-0.062}$, consistent with the {\it Planck} result within $1\,\sigma$, while the high-redshift sample gives a lower value of $S_8 = 0.789^{+0.062}_{-0.062}$. The joint constraint of the two redshift bin yields $S_8 = 0.800^{+0.045}_{-0.045}$ (labeled as ``volume 2 bin joint"). These results show excellent agreement with our flux-limited sample results at low redshifts. The $S_8$ constraints from different samples at low redshifts (Figure \ref{fig:corner_vol}, bottom-left) are consistent at $1\,\sigma$ level.

\textbf{(10) Effect of parameter priors.} We find that the constraints on $\Omega_m$ and $h$ show dependence on the adopted priors. A previous study by \cite{2023_Alonso_quaia_structure_growth} examined $h$ priors based on the \textit{Planck} and supernova data, and found that the prior on $h$ has only a minor impact on the constraint of $S_8$. In contrast, due to the degeneracy between $\Omega_m$ and $\sigma_8$, a higher value of $\Omega_m$ leads to a lower inferred $\sigma_8$. Consequently, unlike the combined parameter $S_8$, the $\sigma_8$ values derived in this work are more sensitive to the accuracy of the BAO measurements from \cite{2017_Alam_BAO_prior} and \cite{2020_Gil_Marin_BAO_prior}.

\textbf{(11) Advantages of high-redshift data.} As shown in Figure \ref{fig:corner}, the high-redshift data are particularly effective at breaking the $\sigma_8$-$\Omega_m$ degeneracy, improving the efficiency of parameter constraints by 12\%. This effectiveness primarily arises from two key factors: the larger cosmological volume probed by high-redshift observations and the increased number of modes captured in the linear regime of the power spectrum. Thus, when future data can address the completeness limitations and the foreground contamination, high-redshift measurements will become more advantageous than their low-redshift counterparts.

\begin{figure*}
    \centering
    \includegraphics[width=1.0\linewidth]{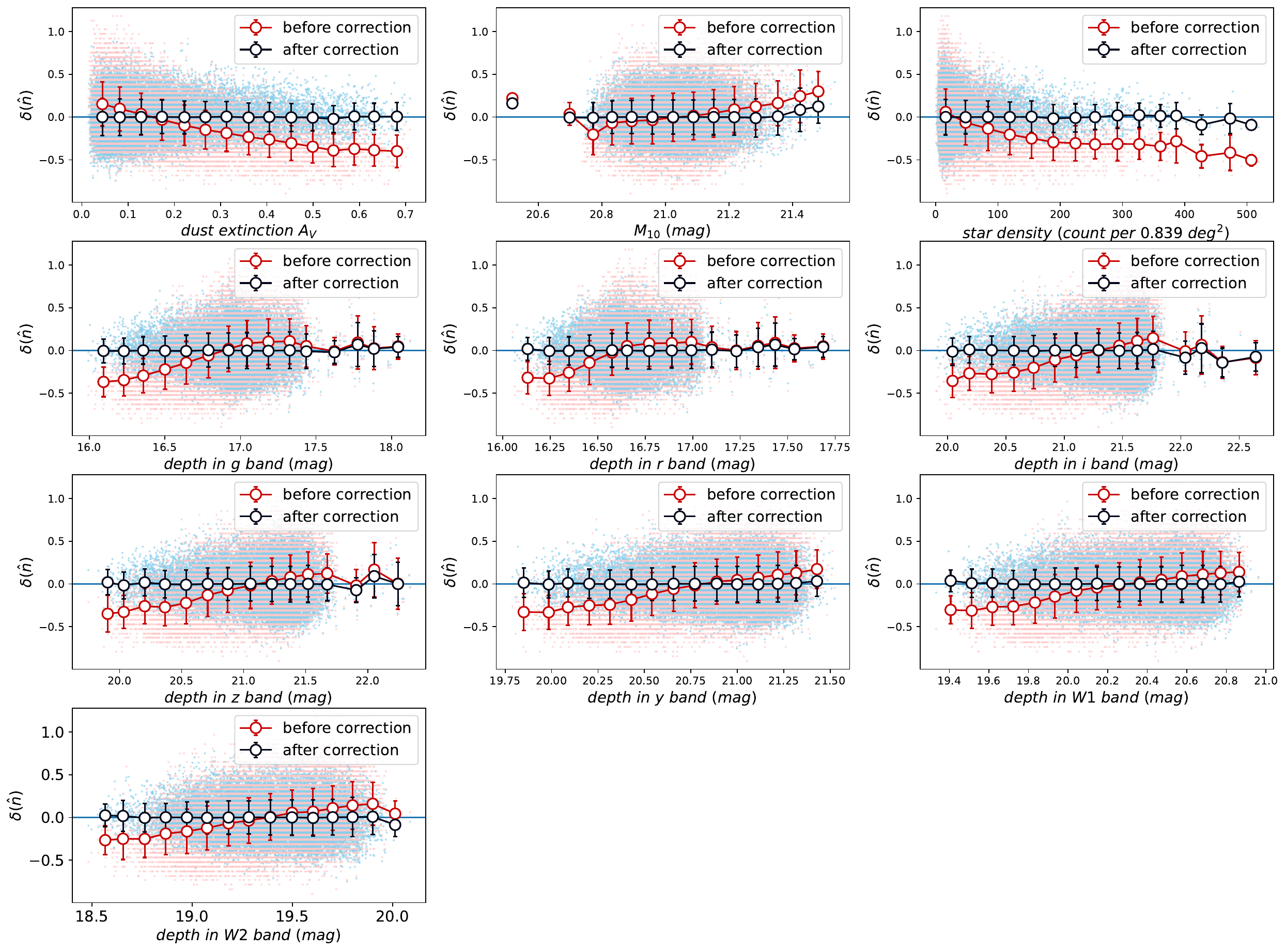}
    \caption{Dependence of quasar overdensity $\delta(\mathbf{\hat{n}})$ on observational systematics before and after correction for the fiducial ``2 bin 1st" case. Each panel displays the relationship between $\delta(\mathbf{\hat{n}})$ and a distinct systematic template: dust extinction, stellar density, and survey depth across multiple photometric bands. The pink and blue points represent the pixel-level overdensities before and after correction, respectively, while the red and black points with error bars show the corresponding binned means and standard deviations. Our correction procedure significantly reduces these systematic correlations, demonstrating that the selection function effectively mitigates the observational biases across the celestial sphere.}
    \label{fig:correlation}
\end{figure*}

\begin{figure*}
    \centering
    \includegraphics[width=1.0\linewidth]{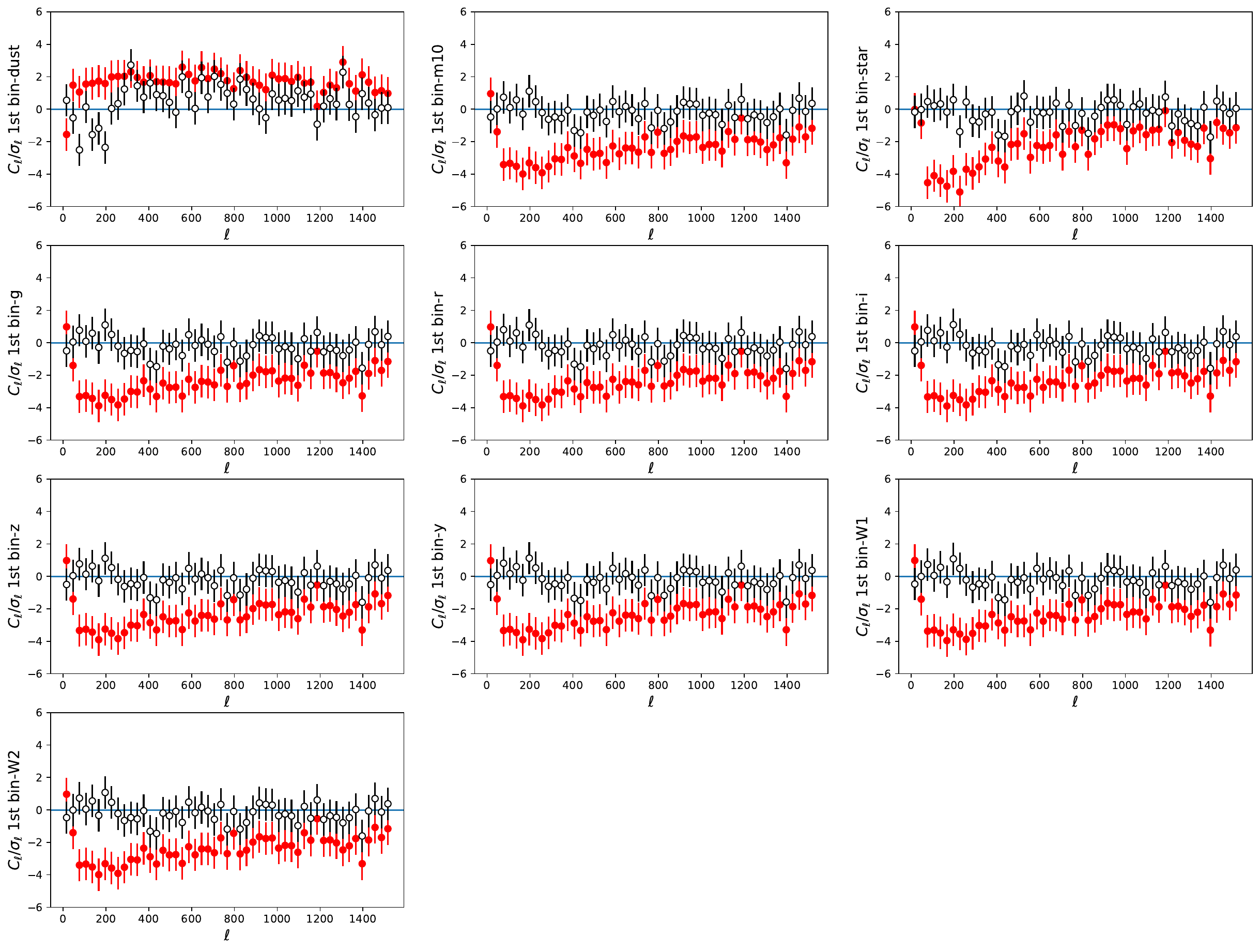}
    \caption{Cross angular power spectra between the quasar overdensity and various systematic templates for the fiducial ``2 bin 1st" case. Each panel corresponds to a distinct systematic map: dust extinction, stellar density, and survey depth in the optical and infrared bands. The red points with error bars represent the cross-power spectra before correction, while the black open circles with error bars show the results after applying the selection function correction. Post-correction, the cross spectra are statistically consistent with zero across all multipoles, demonstrating that the systematic effects have been effectively mitigated.}
    \label{fig:cross_tmp}
\end{figure*}

\textbf{(12) Photo-z uncertainties.}
We perform an additional test for the low-redshift bin (``2 bin 1st") and the high-redshift bin (``2 bin 2nd") by incorporating photometric redshift uncertainties. Following \citet{2020_Nicola_HSC_photoz_uncertainties}, we model the true redshift distribution $p(z)$ using a parameterized form:
\begin{equation}
p(z) \propto \hat{p}\left(z_c+\left(1+z_{w}\right)\left(z-z_c\right)+\Delta z\right)
\end{equation}

where $\hat{p}(z)$ is the estimated redshift distribution, and $p(z)$ is the underlying true redshift distribution. The parameter $z_c$ corresponds to the peak location of $\hat{p}(z)$. The free parameters $z_w$ and $\Delta z$ control the width and the peak position of the redshift distribution, respectively.

We adopt Gaussian priors of $\mathcal{N}(0, 0.1^2)$ for both $z_w$ and $\Delta z$, effectively allowing for a shift in the redshift distribution by up to $\sim 0.1z$ and a $10\%$ change in its width. The resulting constraints are $\Delta z = -0.034^{+0.089}_{-0.091}, z_w = 0.005^{+0.098}_{-0.097}$ for the low-redshift bin and $\Delta z = -0.004^{+0.098}_{-0.099}, z_w = -0.005^{+0.103}_{-0.101}$ for the high-redshift bin. We find that the constraints on $S_8$ for both the low- and high-redshift bins remain robust under this modeling of photometric redshift distribution uncertainties ($S_8 = 0.845^{+0.064}_{-0.063}$ for the low-redshift bin and $S_8 = 0.719^{+0.063}_{-0.061}$ for the high-redshift bin). There is no significant shift in the central values of $S_8$ for either bin.

\textbf{(13) Cross-correlation analysis between low-z and high-z redshift bins.} Figure \ref{fig:cross} shows the cross angular power spectrum between the first and second redshift bins. The measured spectrum shows good agreement with theoretical predictions, confirming the effectiveness of our systematic error corrections in the data processing. The observed non-zero cross-correlation on large scales is attributed to two factors: (1) the overlap in the calibrated redshift distributions between the low- and high-redshift bins, and (2) a potential weak lensing effect, where sources in the high-redshift bin are lensed by the low-redshift bin foreground (though this contribution is expected to be minimal).

\textbf{(14) Cross-correlation analysis between quasar overdensity and systematic templates.}
Figure~\ref{fig:correlation} shows the correlation between quasar overdensity and various systematic templates, using only pixels where the selection function is greater than 0.5. Before applying the selection function correction, $\delta(\mathbf{\hat n})$ exhibits clear trends with the systematics. After correction, the mean values are consistent with zero. The correction procedure substantially reduces the correlations, demonstrating that the selection function effectively mitigates the impact of systematics across the sky. Similarly, we compute the cross angular power spectra between the quasar overdensity and the systematic templates, as shown in Figure~\ref{fig:cross_tmp}. Before correction, the cross spectra exhibit significant non-zero power, indicating residual contamination. After applying the correction, the cross spectra are statistically consistent with zero. These tests confirm that our selection function reliably removes biases in the power spectrum measurements caused by the potential observational systematics.

\textbf{(15) Sensitivity of $S_8$ to the masked regions defined by the selection function.} We further examine how different masks affect the $S_8$ measurements. Specifically, we apply masks defined by $S(\hat{\mathbf{n}}) > 0.4$ and $S(\hat{\mathbf{n}}) > 0.6$ to the two redshift bin cases, resulting in four tests in total. The results are summarized in Table \ref{tab:prior and best fit} and Figure \ref{fig:treeplot}. For the mask with $S(\hat{\mathbf{n}}) > 0.4$, we obtain $S_8 = 0.837^{+0.059}_{-0.058}$ and $0.720^{+0.055}_{-0.054}$ for the low- and high-redshift bins, respectively. For the mask with $S(\hat{\mathbf{n}}) > 0.6$, we find $S_8 = 0.846^{+0.061}_{-0.056}$ and $0.736^{+0.059}_{-0.055}$ for the low- and high-redshift bins, respectively. These results suggest that the choice of mask has only a minor impact on the constraints of $S_8$.

\begin{table*}
	\newcolumntype{l}{>{\raggedright}p{0.2 \textwidth}}
	\newcolumntype{c}{>{\centering}p{0.08\textwidth}}
    \newcolumntype{d}{>{\centering}p{0.1\textwidth}}
	\newcolumntype{r}{>{\centering\arraybackslash}p{0.1\textwidth}}
    \centering
	   \caption{Comparison of Bias Models Using Fit Indices and Posterior Predictive p-Values.}
       \label{tab:AICBIC} 
	   \renewcommand\arraystretch{1.0}
            \rowcolors{2}{gray!10}{white} 
            \begin{tabular}{lcccdcdr}
			\toprule
            \toprule
                 Stat.  & $\chi^2$ &$\chi^2_\nu$ & AIC & AIC-$\rm AIC_{min}$ & BIC &  BIC-$\rm BIC_{min}$  & PPC p-value \\ \midrule
          2 bin 1st & 28.773&0.959&36.773&0&42.377&0&0.531\\
          2 bin 1st constant bias&29.636& 0.988&37.636 &0.863& 43.241&0.864&0.515\\
          2 bin 1st steep bias&28.802 & 0.960 &36.802 &0.029&42.407 &0.030&0.534\\
          2 bin 1st free bias&28.382 & 0.946 & 40.382 &3.609& 48.789 &6.412& 0.553\\
          2 bin 2nd &36.645 &0.679&44.645&0.335&52.601&0.335&0.969\\
          2 bin 2nd constant bias&36.721& 0.680&44.721&0.411&52.677&0.411&0.966\\
          2 bin 2nd steep bias& 36.310&0.672&44.310&0&52.266&0&0.959\\
          2 bin 2nd free bias&36.665& 0.679&48.665&4.355&60.599&8.333&0.967\\
   
			\bottomrule
	   \end{tabular}
\end{table*}

\textbf{(16) Effect of different bias models.} To further test the robustness of our conclusions with respect to the bias model, we repeated the analysis using three different bias parameterizations. The first is a constant-bias model $b(z)=b_g$. The second model is $b(z)=b_g\left[1+\left(\frac{1+z}{2.5}\right)^5\right]$, which has a steep power-law index of 5. The third model is a fully free, three-parameter functional form $b(z)=a(1+z)^\alpha+b$. The first two models follow the tests performed in \cite{2023_Alonso_quaia_structure_growth}. The resulting constraints are summarized in Table \ref{tab:prior and best fit}. We note that the fully free bias model does not allow all three bias parameters ($a$, $b$, and $\alpha$) to be independently constrained. As a result, their posterior distributions are non-Gaussian, and therefore we do not list them in Table \ref{tab:prior and best fit}. However, the cosmological parameters $\Omega_m$, $\sigma_8$, $h$, and $S_8$ remain well constrained. 

For the first bin, labeled as ``2 bin 1st constant bias”, ``2 bin 1st steep bias”, and ``2 bin 1st free bias”, we obtain $S_8 = 0.852^{+0.057}_{-0.058}$, $S_8 = 0.826^{+0.060}_{-0.054}$, and $S_8 = 0.842^{+0.060}_{-0.054}$. For the second bin, labeled as ``2 bin 2nd constant bias”, ``2 bin 2nd steep bias”, and ``2 bin 2nd free bias”, the corresponding results are $S_8 = 0.757^{+0.058}_{-0.057}$, $S_8 = 0.681^{+0.054}_{-0.051}$, and $S_8 = 0.742^{+0.057}_{-0.059}$. These results indicate that (1) the high-redshift bin is more sensitive to the assumed bias model; (2) the constant-bias model tends to produce a larger value of $S_8$, whereas the steep bias model yields a smaller $S_8$; and (3) the free bias model gives an $S_8$ value that lies between these two extreme cases (which are known to deviate significantly from observations) and is close to the result obtained with our fiducial bias model of Equation \eqref{bias}. In a nutshell, although different bias models do shift the inferred value of $S_8$, particularly in the high-redshift bin, they do not alter our main conclusions: the low-redshift result remains consistent with the Planck 2018 constraint, while the high-redshift bin still shows a lower $S_8$ value.

We compare the bias models using the goodness-of-fit statistics and information criteria summarized in Table \ref{tab:AICBIC}. For each redshift bin, we report the minimum $\chi^2$, the reduced $\chi^2$, the Akaike Information Criterion (AIC), the Bayesian Information Criterion (BIC), and the posterior predictive check (PPC) $p$-value. In the PPC analysis, the $\chi^2$ statistic is adopted as the test quantity, and 1000 samples from the MCMC chains are drawn to generate the posterior predictive distribution. In the first redshift bin, all bias models provide comparably good fits, with reduced $\chi^2$ values close to unity. The fiducial, constant, and steep bias models yield nearly identical AIC and BIC values, while the free bias model is disfavored ($\Delta\mathrm{AIC}\simeq 3.6$, $\Delta\mathrm{BIC}\simeq 6.4$). The PPC $p$-values lie within the range expected for an acceptable model fit. In the second redshift bin, all models again achieve similar goodness-of-fit, with reduced $\chi^2$ values below unity, which may reflect a conservative covariance estimate. The fiducial, constant, and steep bias models remain statistically indistinguishable, whereas the free bias model is strongly disfavored ($\Delta\mathrm{AIC}\simeq 4.4$, $\Delta\mathrm{BIC}\simeq 8.3$). The PPC $p$-values are correspondingly high.

\section{Conclusion} \label{sec:conclusion}
This study utilizes the CatNorth catalog of 1.5 million quasar candidates \citep{Fu_2024_CatNorth}—one of the most extensive and complete quasar candidate catalogs to date—to investigate the $S_8$ tension. This is achieved by applying the machine learning-based selection function to correct for the spatial incompleteness and by incorporating the full-sky {\it Planck} CMB lensing data \citep{2022_Carron_CMB_lensing}.
By dividing the sample into low- and high-redshift bins, we find that the low-redshift measurement, $S_8 = 0.844^{+0.058}_{-0.056}$ (7\% precision), is consistent with the {\it Planck} 2018 CMB anisotropy result of $S_8 = 0.834 \pm 0.016$ \citep{2018_planck_cosmology_parameters} at the $1\,\sigma$ level. In contrast, the high-redshift bin ($z > 1.5$) yields a lower value of $S_8 = 0.724^{+0.058}_{-0.054}$, likely due to incompleteness in the faint, high-redshift quasar population. The additional volume-limited samples yield results consistent with those from the flux-limited sample.

For the $S_8$ tension, quasars serve as a valuable complementary probe independent of CMB anisotropies and galaxy weak lensing. Photometric quasar candidate samples span wide sky areas and encompass large cosmological volumes. When combined with well-defined selection functions and accurate photometric redshift estimates, they can provide precise constraints on cosmological parameters.

Interestingly, recent results from other cosmological probes also indicate a potential alleviation of the $S_8$ tension. For example, the latest measurements from KiDS-Legacy yield a relatively high value of \(S_8 = 0.815^{+0.016}_{-0.021}\), which differs significantly from the earlier KiDS-1000 result. This discrepancy is attributed to the improved redshift calibration, an expanded survey area, and the enhanced image reduction \citep{2025_kids_legacy_cosmologicalconstraintscosmic}. From a joint analysis of galaxy clustering and weak lensing using DESI DR9 data and a minimal bias model, \citet{2023_Xu_S8} find that \(S_8\) exhibits a mild increase with redshift. Combining their results with low-redshift data, they obtain \(S_8 = 0.84 \pm 0.02\), consistent with the \textit{Planck} results. Similarly, \citet{2024_Xu_JingYipeng_S8} analyze the gravitational lensing magnification around BOSS CMASS galaxies and report \(S_8 = 0.816 \pm 0.024\), in agreement with both the \textit{Planck} and WMAP. However, they observe an enhanced magnification signal at small scales (\(r_p < 70\,h^{-1}\mathrm{kpc}\)), possibly indicating baryonic feedback effects. \citet{2025_luo_photometricobjectscosmicwebs_S8} apply the Photometric Objects Around Cosmic Webs (PAC) method to reconstruct galaxy-halo connections and find \(S_8 = 0.8294 \pm 0.0110\) (CMASS with HSC), \(S_8 = 0.8073 \pm 0.0372\) (CMASS with DES), and \(S_8 = 0.8189 \pm 0.0440\) (CMASS with KiDS). Their results suggest that no significant baryonic feedback is required to suppress small-scale clustering.

Looking ahead, the next-generation wide-field telescopes—such as the Vera C. Rubin Observatory's Legacy Survey of Space and Time (LSST) \citep{2019_Ivezic_LSST}, {\it Euclid} \citep{mellier_2024_Euclid}, and the Chinese Space Station Survey Telescope (CSST, covering approximately 17,500 deg$^2$) \citep{CSST_zhan_hu_2011, CSST_zhan_hu_2021}—are poised to transform quasar surveys. These facilities are expected to enable complete quasar samples reaching redshifts up to $z \sim 3$ and increase the sample sizes by up to at least one order of magnitude. By delivering high-precision redshift measurements across broad sky coverage and multiple cosmic epochs, these surveys will be instrumental in uncovering the origin of the $S_8$ tension—whether it arises from the residual systematics or points to new physics.

\begin{acknowledgments}
We thank the anonymous referee for the constructive report, which has greatly helped in revising the paper. We thank Ji Yao for valuable discussions and insightful comments that helped improve this work.

We thank the support of the National Key R\&D Program of China (grant No.2025YFA14101) and the National Science Foundation of China (grant No. 12133001).

This work makes use of data from the European Space Agency (ESA) mission \href{https://www.cosmos.esa.int/gaia}{Gaia}, processed by the \href{https://www.cosmos.esa.int/web/gaia/dpac/consortium}{Gaia Data Processing and Analysis Consortium (DPAC)}. Funding for DPAC has been provided by national institutions, in particular the institutions participating in the Gaia Multilateral Agreement.

This work makes use of data from the \href{https://outerspace.stsci.edu/display/PANSTARRS}{Pan-STARRS1 Surveys (PS1)}. The PS1 and the PS1 public science archive have been made possible through contributions by the Institute for Astronomy at the University of Hawaii; the Pan-STARRS Project Office; the Max Planck Society and its participating institutes, including the Max Planck Institute for Astronomy (Heidelberg) and the Max Planck Institute for Extraterrestrial Physics (Garching); The Johns Hopkins University; Durham University; the University of Edinburgh; Queen's University Belfast; the Harvard-Smithsonian Center for Astrophysics; the Las Cumbres Observatory Global Telescope Network; the National Central University of Taiwan; the Space Telescope Science Institute; the National Aeronautics and Space Administration (NASA) under Grant No. NNX08AR22G issued through the Planetary Science Division of the NASA Science Mission Directorate; the National Science Foundation (Grant No. AST1238877); the University of Maryland; E{\"o}tv{\"os} Lor{\'a}nd University (ELTE); Los Alamos National Laboratory; and the Gordon and Betty Moore Foundation.

This work makes use of data from the \href{https://www.jpl.nasa.gov/missions/wide-field-infrared-survey-explorer-wise}{Wide-field Infrared Survey Explorer (WISE)}, a joint project of the University of California, Los Angeles, and the Jet Propulsion Laboratory/California Institute of Technology, funded by NASA.

This work makes use of data from the \href{https://www.esa.int/Science_Exploration/Space_Science/Planck}{Planck satellite}, a space observatory operated by the European Space Agency (ESA) designed to map the anisotropies of the cosmic microwave background at microwave and infrared frequencies with high sensitivity and angular resolution. We thank the Planck Collaboration for their efforts in acquiring, processing, and publicly releasing the data. The Planck data used in this work were obtained from the \href{https://www.cosmos.esa.int/web/planck}{ESA Planck Legacy Archive}.

Funding for the \href{https://www.sdss4.org/}{Sloan Digital Sky 
Survey IV} has been provided by the Alfred P. Sloan Foundation, the U.S. Department of Energy Office of Science, and the Participating Institutions. 

SDSS-IV acknowledges support and 
resources from the Center for High 
Performance Computing  at the 
University of Utah. The SDSS 
website is www.sdss4.org.

SDSS-IV is managed by the 
Astrophysical Research Consortium 
for the Participating Institutions 
of the SDSS Collaboration including 
the Brazilian Participation Group, 
the Carnegie Institution for Science, 
Carnegie Mellon University, Center for 
Astrophysics | Harvard \& 
Smithsonian, the Chilean Participation 
Group, the French Participation Group, 
Instituto de Astrof\'isica de 
Canarias, The Johns Hopkins 
University, Kavli Institute for the 
Physics and Mathematics of the 
Universe (IPMU) / University of 
Tokyo, the Korean Participation Group, 
Lawrence Berkeley National Laboratory, 
Leibniz Institut f\"ur Astrophysik 
Potsdam (AIP),  Max-Planck-Institut 
f\"ur Astronomie (MPIA Heidelberg), 
Max-Planck-Institut f\"ur 
Astrophysik (MPA Garching), 
Max-Planck-Institut f\"ur 
Extraterrestrische Physik (MPE), 
National Astronomical Observatories of 
China, New Mexico State University, 
New York University, University of 
Notre Dame, Observat\'ario 
Nacional / MCTI, The Ohio State 
University, Pennsylvania State 
University, Shanghai 
Astronomical Observatory, United 
Kingdom Participation Group, 
Universidad Nacional Aut\'onoma 
de M\'exico, University of Arizona, 
University of Colorado Boulder, 
University of Oxford, University of 
Portsmouth, University of Utah, 
University of Virginia, University 
of Washington, University of 
Wisconsin, Vanderbilt University, 
and Yale University.
\end{acknowledgments}


\facilities{Gaia, PS1, WISE, Planck, Sloan}

\software{\texttt{astropy} \citep{astropy}, \texttt{healpy} and \texttt{HEALPix} \citep{2005_Gorski_healpix}, \texttt{pyccl} \citep{Chisari_2019_PYCCL}, \texttt{pymaster} and \texttt{NaMaster} \citep{2019_Alonso_NaMaster}, \texttt{emcee} \citep{emcee}, \texttt{corner} \citep{corner}, \texttt{getdist} \citep{getdist}, \texttt{numpy} \citep{numpy}, \texttt{scipy} \citep{scipy}, \texttt{matplotlib} \citep{matplotlib}, \texttt{pytorch} \citep{pytorch}}

\appendix
\renewcommand{\thefigure}{A\arabic{figure}}
\setcounter{figure}{0}

\section{Selection function}
\label{appendix: selection function}

\begin{figure*}
    \centering
    \includegraphics[width=0.32\linewidth]{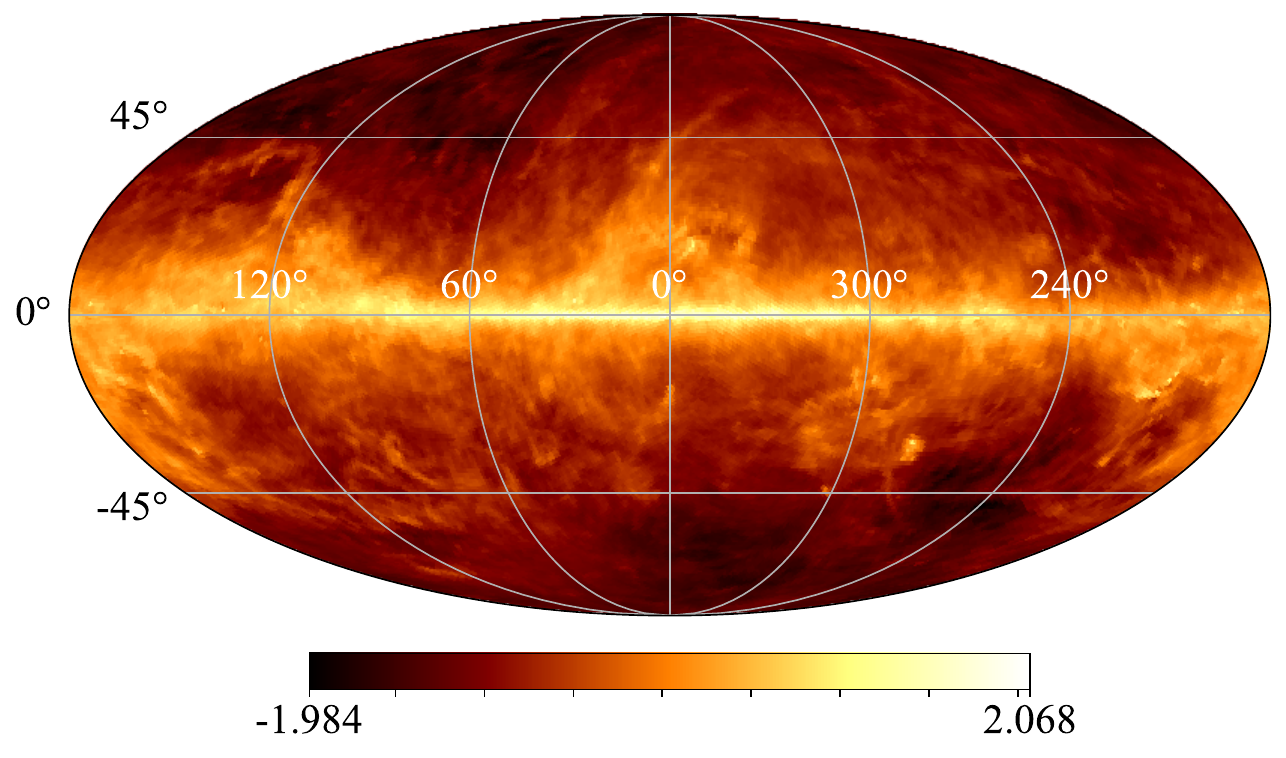}
    \\
    \includegraphics[width=0.32\linewidth]{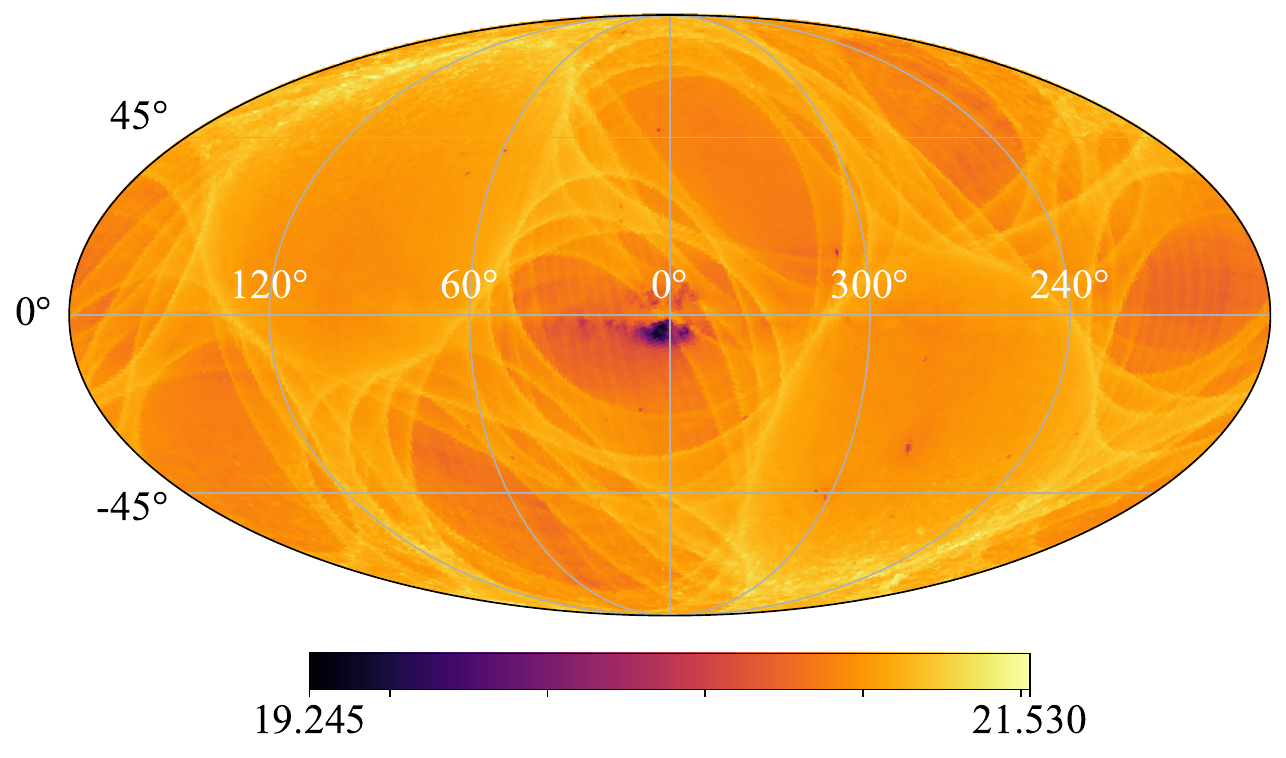}
    \includegraphics[width=0.32\linewidth]{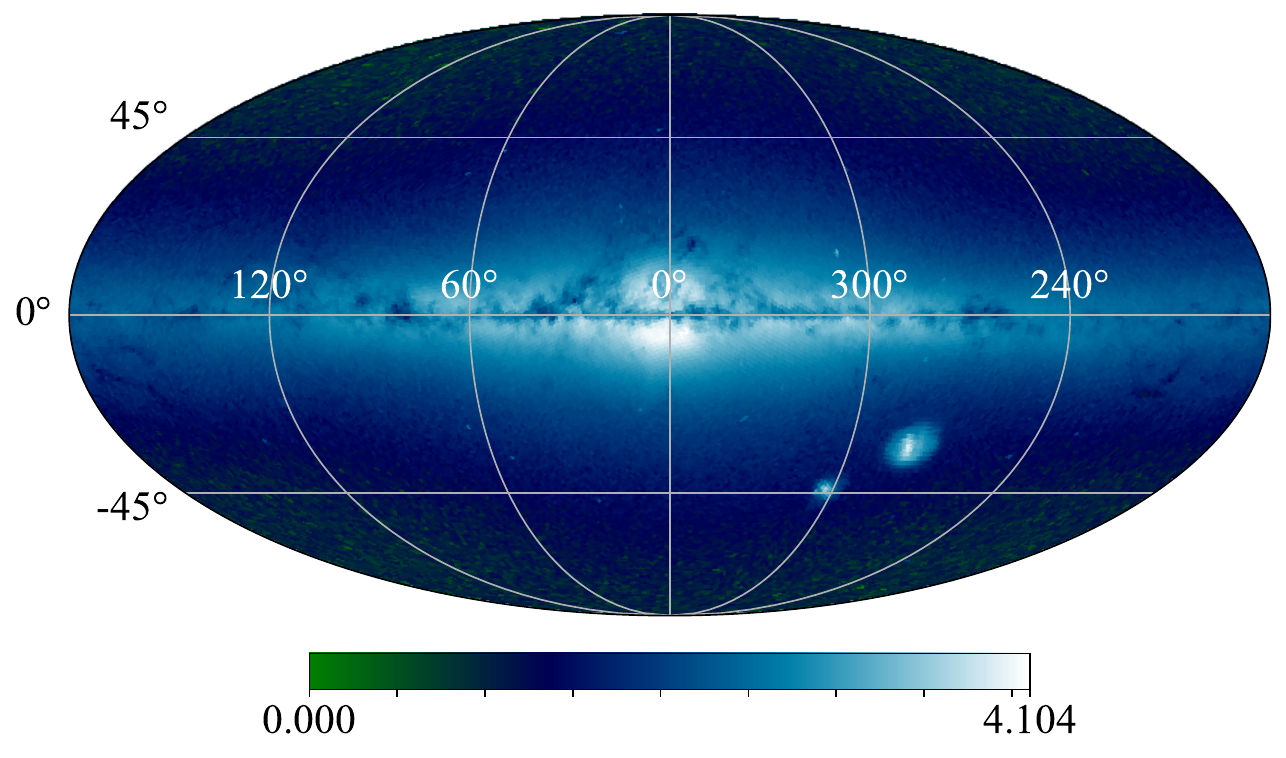}
    \\
    \includegraphics[width=0.32\linewidth]{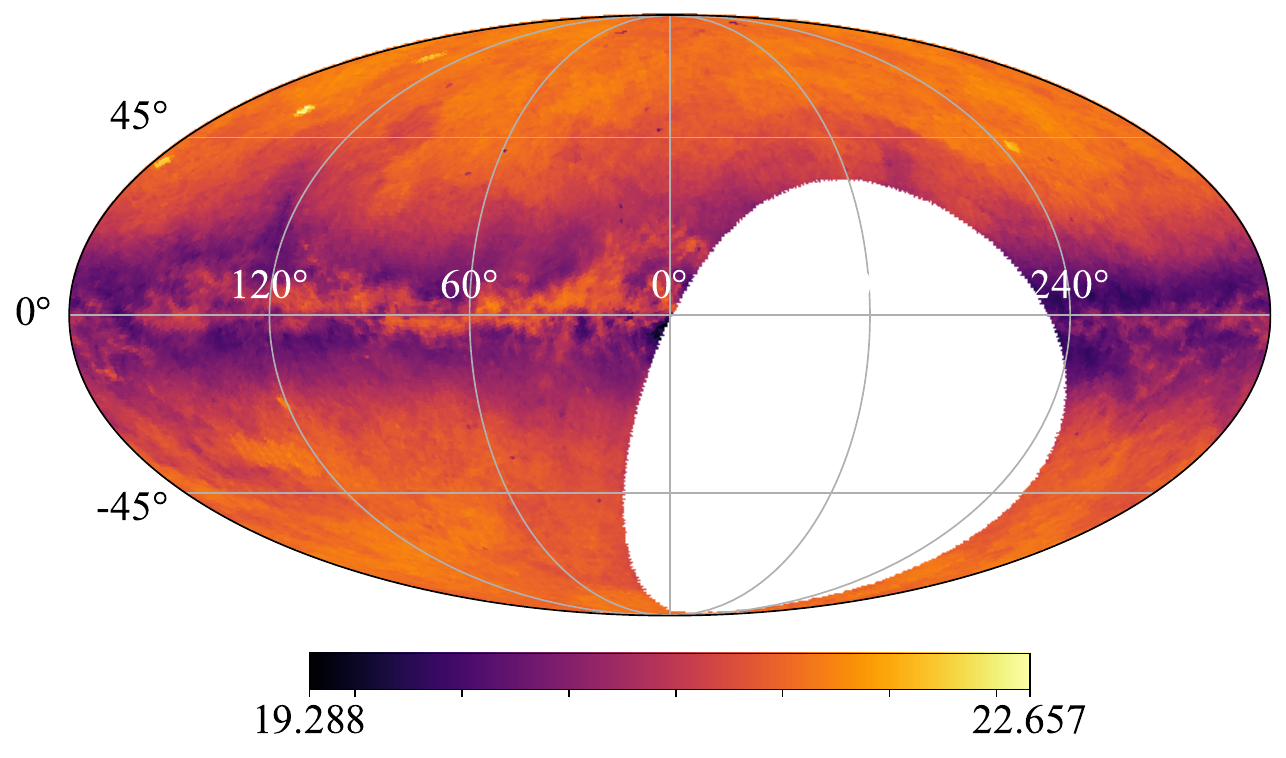}
    \includegraphics[width=0.32\linewidth]{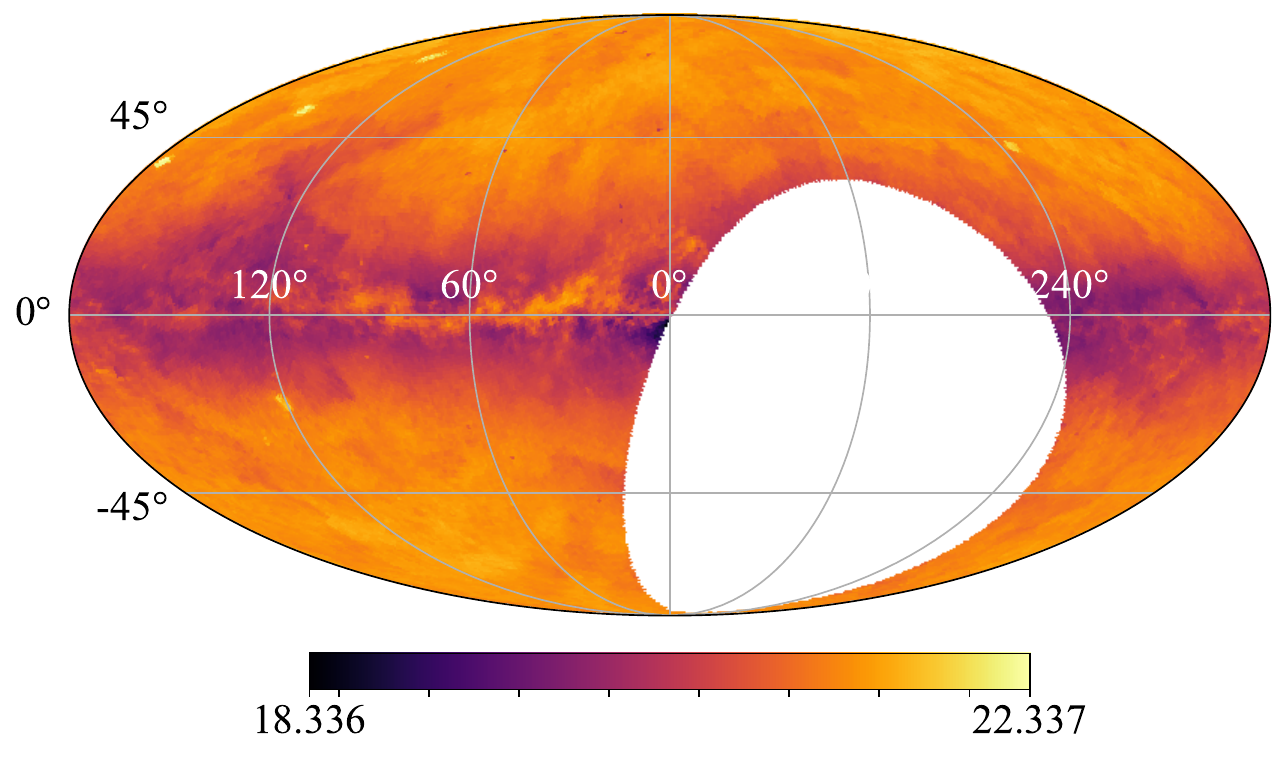}
    \includegraphics[width=0.32\linewidth]{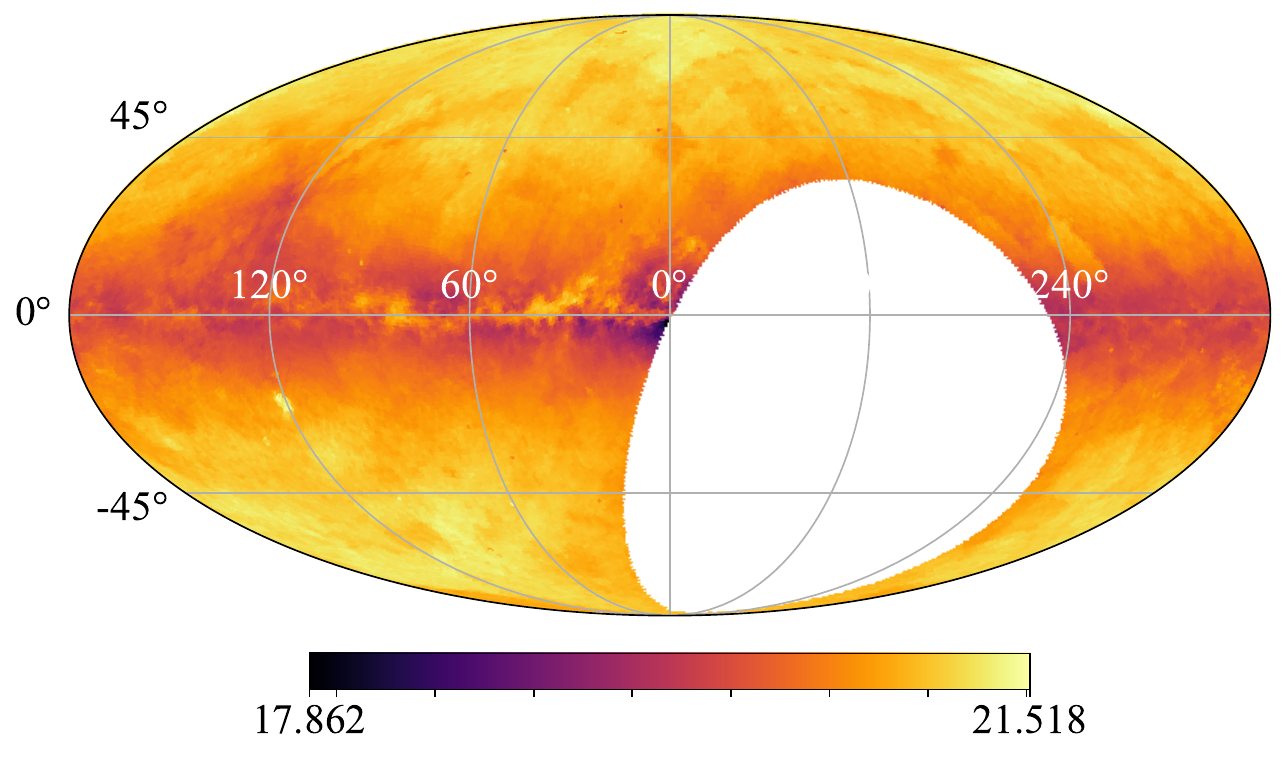}
    \includegraphics[width=0.32\linewidth]{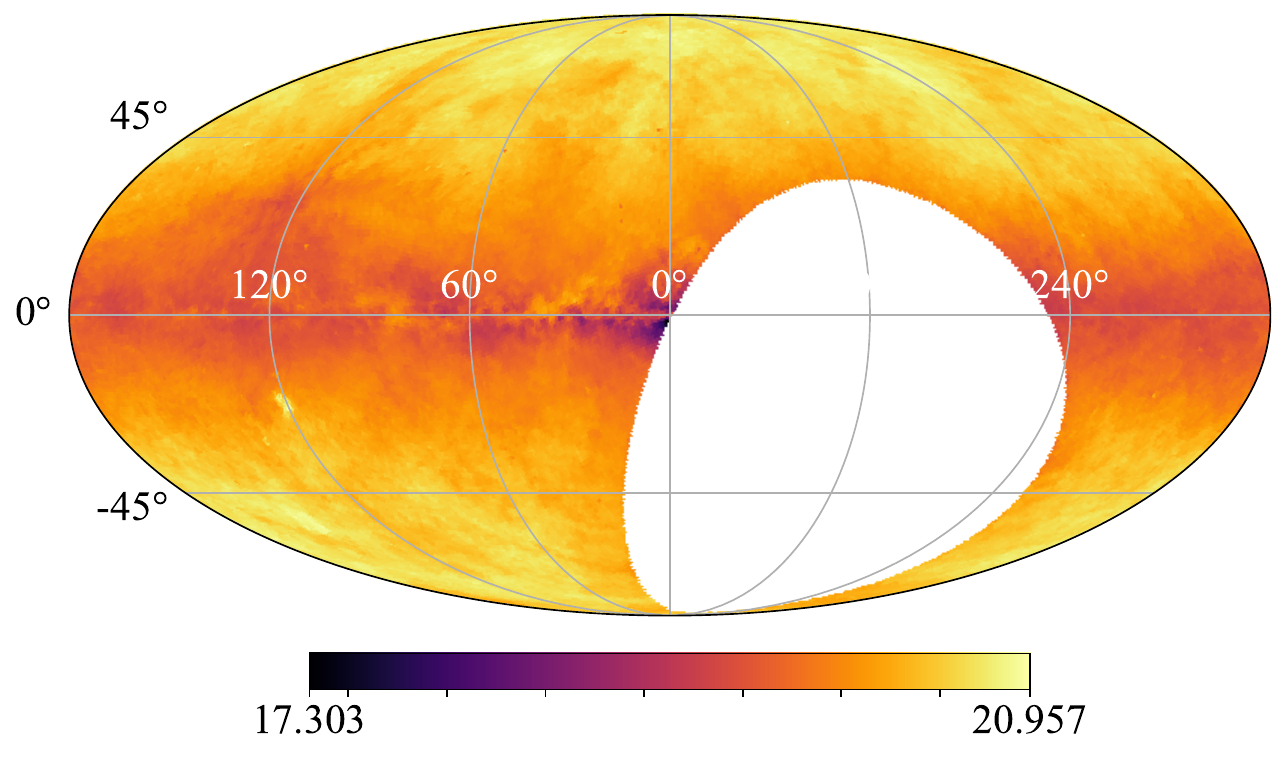}
    \includegraphics[width=0.32\linewidth]{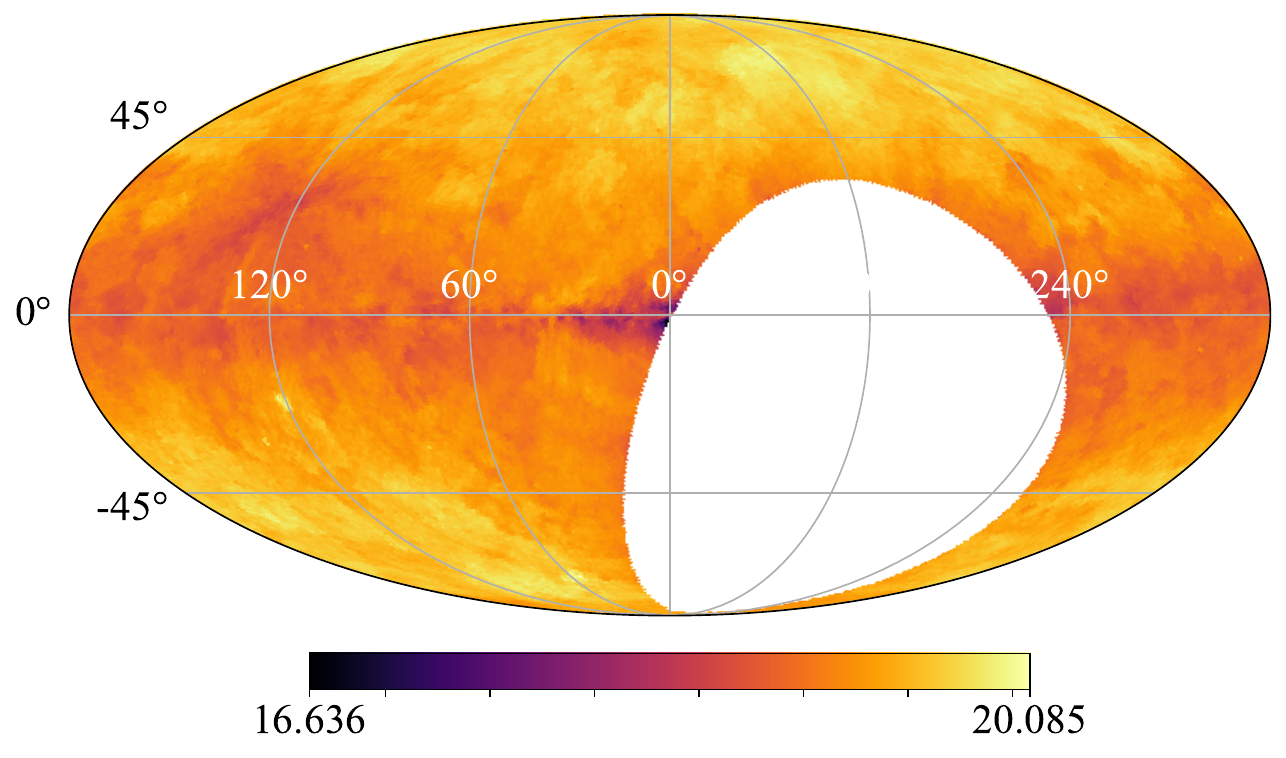}
    \\
    \includegraphics[width=0.32\linewidth]{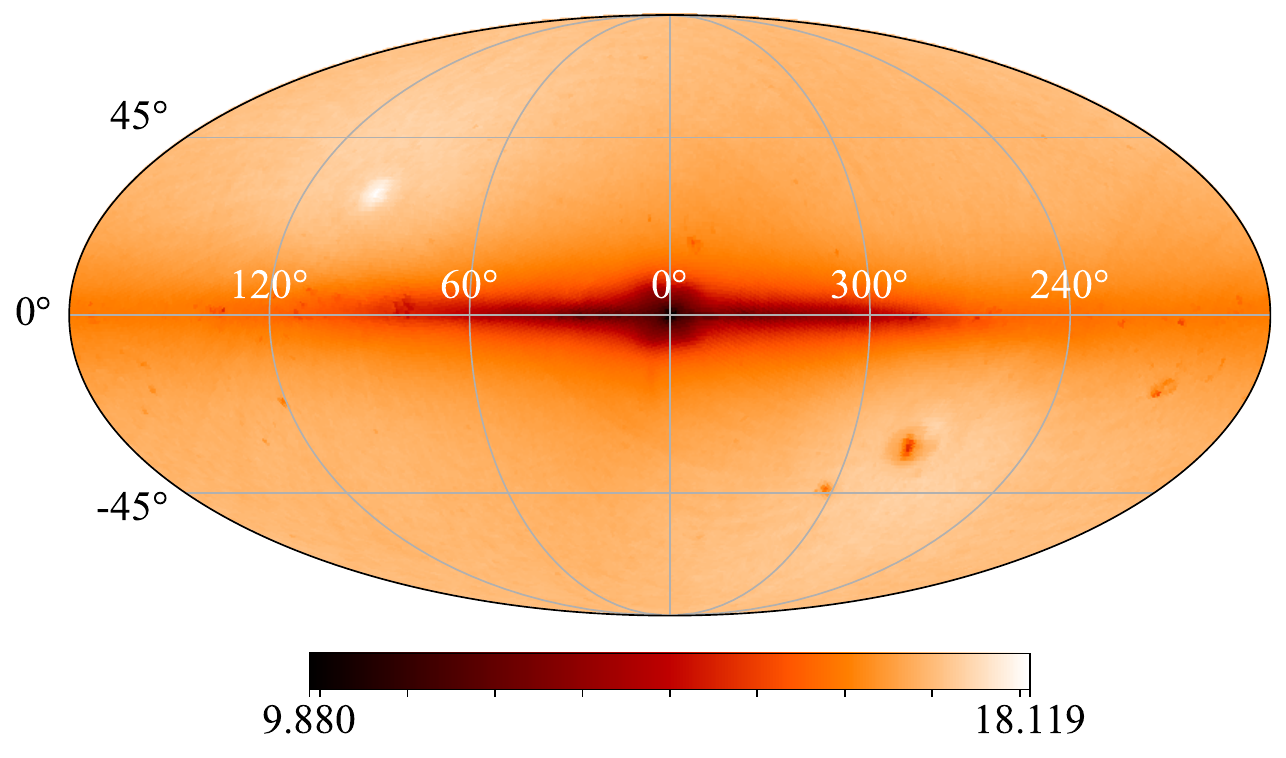}
    \includegraphics[width=0.32\linewidth]{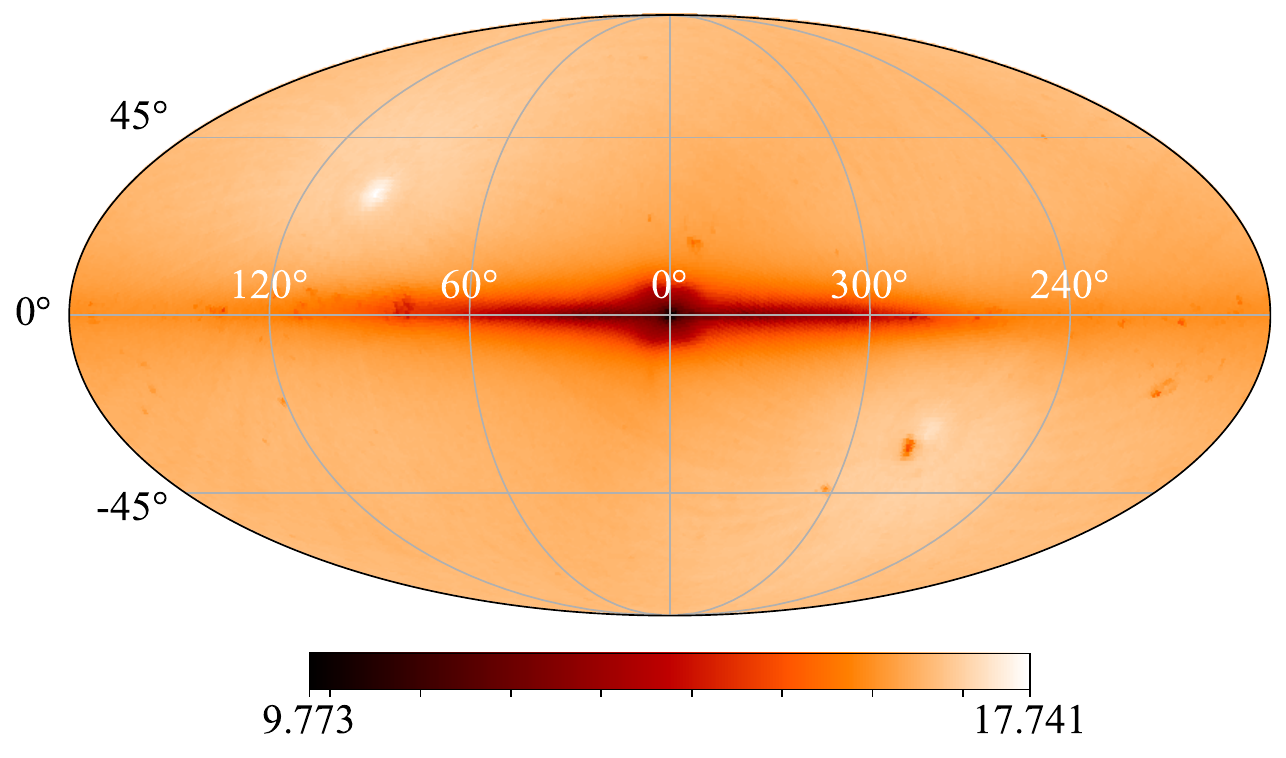}

    \caption{Systematics templates used to construct the selection function, including the dust extinction map $\log_{10}(A_V)$ (first row), Gaia $M_{10}$ map (second row, left), Gaia star density distribution $\log_{10}(n_\star)$ (second row, right), the Pan-STARRS1 median magnitudes in the $\texttt{g}$ (third row, left), $\texttt{r}$ (third row, middle), $\texttt{i}$ (third row, right), $\texttt{z}$ (fourth row, left), and $\texttt{y}$ (fourth row, right) bands, the CatWISE2020 median magnitudes in the $\texttt{W1}$ (fifth row, left) and $\texttt{W2}$ (fifth row, right) bands, serving as proxies for the combined effects of the telescope's scanning law and source crowding.}
    \label{fig:templates}
\end{figure*}

We assume that the spatial distribution of quasars consists of two components: a uniform field $ n_u(\mathbf{\hat{n}}) $ in the sky direction, represented by the unit vector $ \mathbf{\hat{n}} $, which is based on the cosmological principle and corresponds to zeroth-order cosmology; and a fluctuation field $ \delta(\mathbf{\hat{n}}) $, which includes both the linear and non-linear large-scale structures of higher-order cosmology. Thus, in principle, the observed number of sources, $ n_{\text{obs}}(\mathbf{\hat{n}}) $, can be expressed as:  
\begin{equation}
n_{\text{obs}}(\mathbf{\hat{n}})  =  S(\mathbf{\hat{n}})\left[n_u(\mathbf{\hat{n}})  + n_\delta(\mathbf{\hat{n}}) + n_c(\mathbf{\hat{n}}) \right],
\label{eq:model0}
\end{equation}
where \( S(\mathbf{\hat{n}}) \) is a factor that represents the systematic bias, or the so-called selection function, and $ n_c(\mathbf{\hat{n}}) $ is the number of contamination sources incorrectly identified as quasars. In most cases, $ n_c(\mathbf{\hat{n}}) $ originates from nearby galaxies or globular clusters. For example, in regions near M31 and M33, the number density of quasar candidates is high; however, most of these sources are likely stars rather than genuine quasars. Previous analysis of the Quaia quasar candidate catalog revealed significant issues in the LMC and SMC regions when performing Gaussian Process regression to construct the selection function \citep{2024_Storey-Fisher_quaia_catalog}. For cosmological studies, these regions pose challenges because observing and identifying galaxies or quasars behind nearby galaxies or globular clusters is difficult. This leads to the loss of cosmological fluctuation information and reduces the precision of the selection function. We suggest that before constructing the selection function, regions with a significant excess of $ n_c(\mathbf{\hat{n}}) $ should be masked. In this work, we mask the M31 and M33 regions, which is equivalent to setting $ n_c(\mathbf{\hat{n}}) = 0 $. Therefore, we can rewrite the equation \eqref{eq:model0} as:  
\begin{equation}
 n_{obs}(\mathbf{\hat{n}})  =  S(\mathbf{\hat{n}})\left[n_u(\mathbf{\hat{n}})  + n_\delta(\mathbf{\hat{n}})\right],
 \label{eq:model1}
\end{equation}
To construct CatNorth, a cross-match between Gaia DR3, PanSTARRS1, and CatWISE2020 is required. Therefore, CatNorth inherits all the selection effects from Gaia, PanSTARRS1, and CatWISE2020, and its depth is determined by the shallowest of the three surveys. All of these surveys are affected by the foreground dust extinction in the Milky Way, as well as by the detection efficiency of the telescopes, including scanning patterns (which are especially crucial for Gaia) and crowding effects, which arise from the difficulty of distinguishing blended sources during the PSF fitting.

Galactic extinction can lead to an underestimation of the intrinsic brightness of observed objects, which is particularly significant in the optical bands. As a result, many objects become fainter and harder to detect with telescopes. Extinction affects the blue light more strongly, leading to the reddening of objects. If the extinction correction is not performed accurately, it can alter the colors of sources, impacting classifications based on color cuts or other color-dependent methods. This may introduce contamination (mainly from stars) into the sample or cause some quasars to be incorrectly excluded from catalogs. Galactic extinction varies significantly across the sky, being stronger near the Galactic plane and weaker at higher Galactic latitudes.
Figure \ref{fig:templates} shows the extinction map used in this work, which is a modified version of the dust map from \cite{2023_Chiang_dustmap}, as presented in \cite{2024_Storey-Fisher_quaia_catalog}.

\begin{figure}
    \centering    \includegraphics[width=0.9\linewidth]{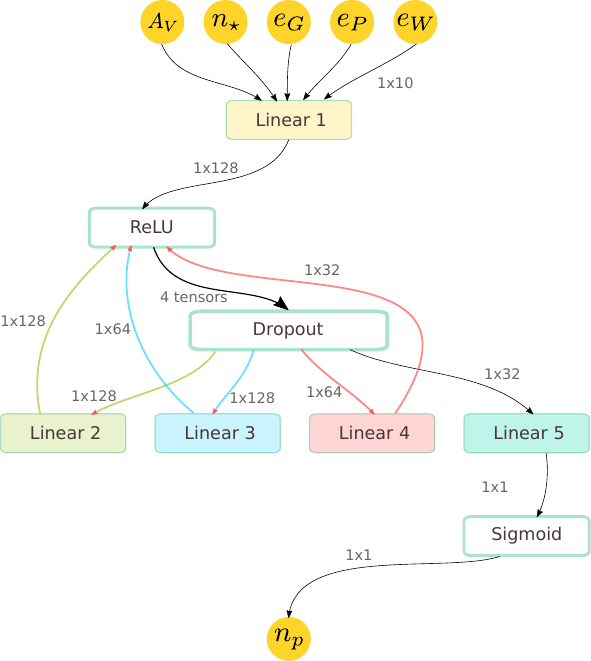}
    \caption{The computational graphs of the MLP neural networks used in this paper. The neural network we used has 10 input features, including dust extinction $A_V$, star density $n_\star$, Gaia $M_{10}$, and Pan-STARRS1 depth in five bands (\texttt{g},\texttt{r},\texttt{i},\texttt{z},\texttt{y}), as well as the CatWISE2020 depth map in two bands (\texttt{W1}, \texttt{W2}). All of these input features are normalized to the range [0, 1] before training. The network consists of five hidden linear layers with shapes of (10, 128), (128, 128), (128, 64), (64, 32), and (32, 1). A ReLU activation function is applied between each layer to improve the model's ability to fit non-linear features, and a Dropout layer is used to prevent overfitting.}
    \label{fig:fnn}
\end{figure}

As a space telescope, Gaia is largely unaffected by seeing, atmospheric extinction, and weather-related issues. However, Gaia follows a specific scanning pattern to optimize the survey efficiency, which means that it must prioritize certain sources while discarding others when there are too many sources in the field of view. \cite{2023_Cantat_Gaudin_Gaia_selection_function} found that the median magnitude \( M_{10} \) in a patch of the sky, for cataloged sources with \(\texttt{astrometric\_matched\_transits} \leq 10\), serves as a good proxy for Gaia's scanning pattern, which is shown in the left panel, second row of Figure \ref{fig:templates}. The detection efficiency of Gaia is also affected by star density, as shown in the right panel, second row of Figure \ref{fig:templates}, which is taken from \cite{2024_Storey-Fisher_quaia_catalog}.  
In this work, we use these two maps to represent the detection efficiency of Gaia.

Various observables from catalogs can be used to estimate the detection efficiency of a telescope, such as the magnitude at which the observed luminosity function deviates from an expected power law, the mode of the magnitude distribution, the magnitude of the faintest star in a given area, and the median magnitude. Although the computation is time-consuming and requires processing the entire catalog, the median magnitude map (or other percentiles, such as the 90th percentile) is a more robust quantity for representing the detection efficiency of a telescope \citep{2023_Cantat_Gaudin_Gaia_selection_function}. Therefore, we use the median magnitude as a proxy for the telescope's detection efficiency for both Pan-STARRS1 and CatWISE2020.

The median magnitudes in five bands for Pan-STARRS1 are shown in the third and fourth rows of Figure \ref{fig:templates}. As observed, Pan-STARRS' detection efficiency generally improves with the increasing Galactic latitude, primarily due to the effects of Galactic extinction. Additionally, these maps exhibit patchy patterns, which are caused by varying observational conditions, such as weather, airmass, and seeing during different observing periods of the ground-based survey. 

The median magnitude maps for the CatWISE2020 $\texttt{W1}$ and $\texttt{W2}$ bands are shown in the fifth row of Figure \ref{fig:templates}. It is evident that, compared to Pan-STARRS, WISE is less affected by Galactic extinction. However, the overall distribution still shows higher completeness in the high Galactic latitude regions. Additionally, due to WISE's overlapping field-of-view tiles, areas around the North and South Ecliptic poles experience more visits, resulting in the observation of more faint sources in these regions.

\begin{figure}
    \centering
    \includegraphics[width=1.0\linewidth]{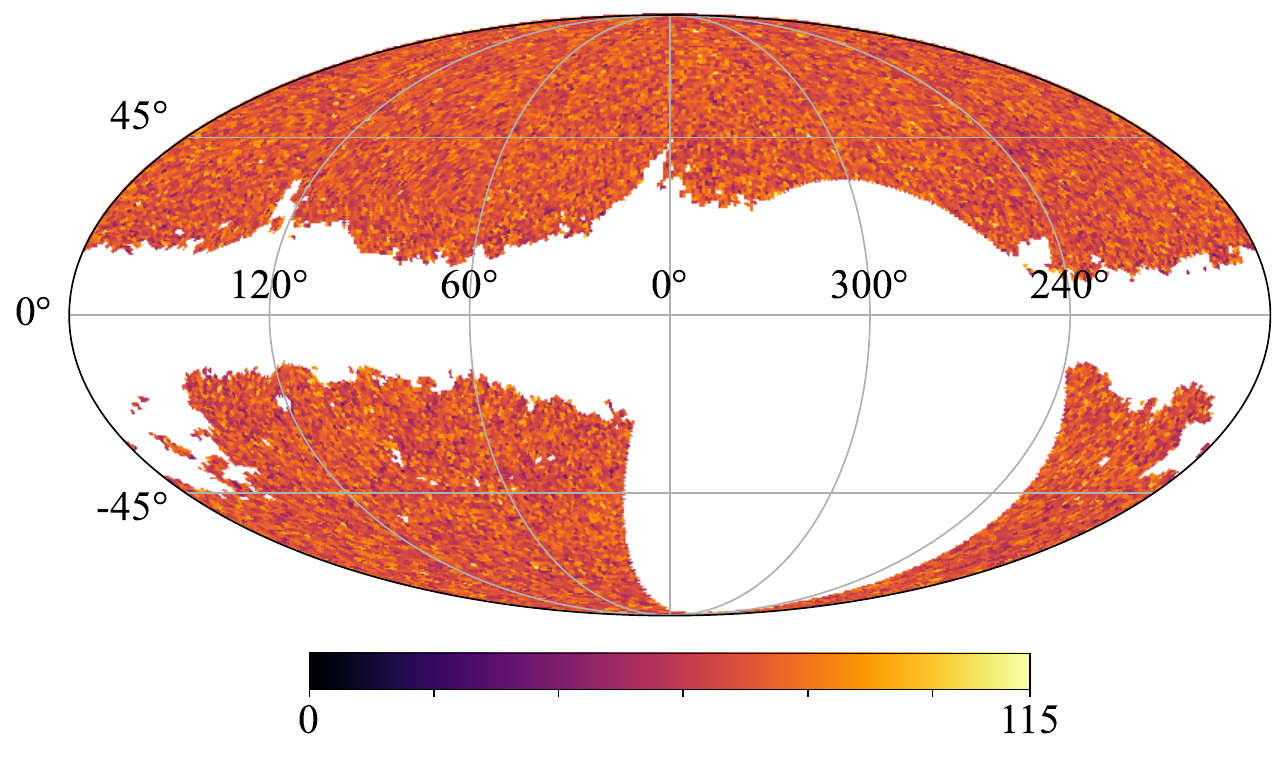}
    \caption{The sky density map of the CatNorth quasar candidate catalog, corrected by the selection function, is presented in Galactic coordinates. The map has a resolution parameter of $N_{\rm side}=64 $. The masked region near the galactic plane satisfies $S(\mathbf{\hat{n}})< 0.5$ to avoid numerical instability when $S(\mathbf{\hat{n}})$ becomes too small. The color bar is expressed in units of counts per 0.839 square degrees.}
    \label{fig:qso_real}
\end{figure}

\begin{figure}
    \centering
    \includegraphics[width=1.0\linewidth]{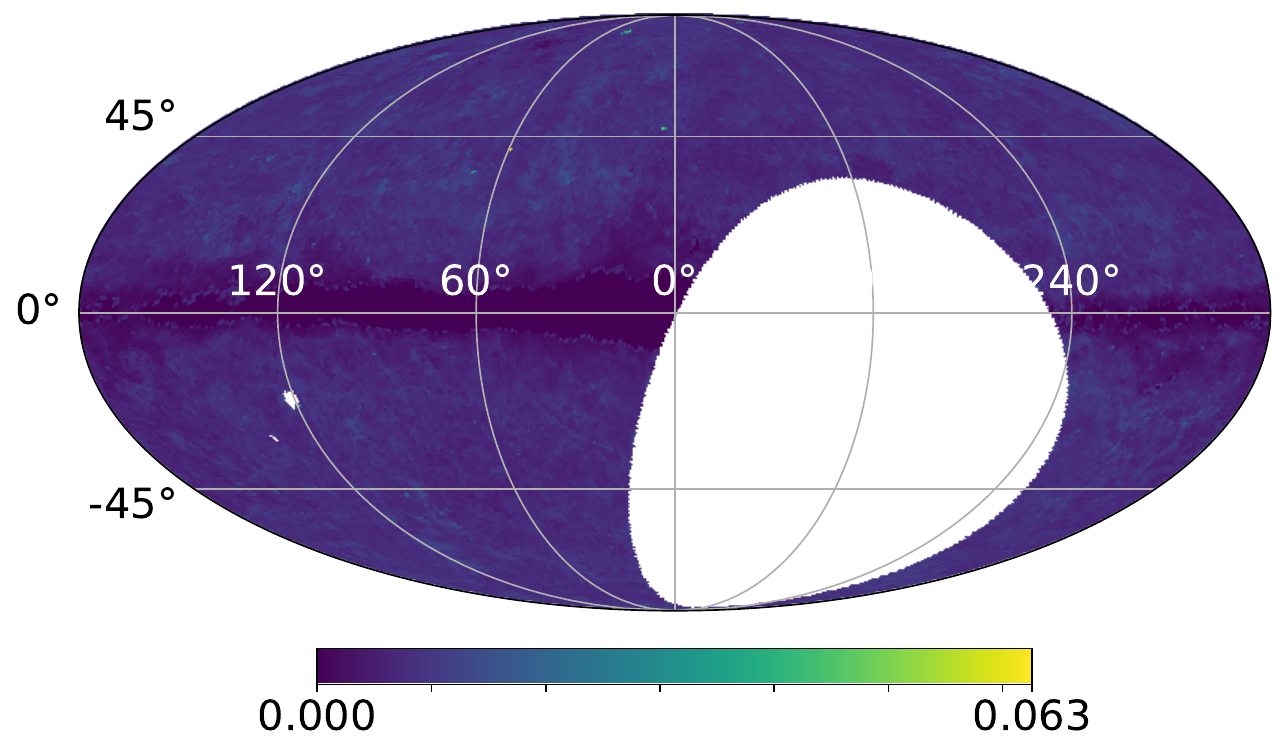}
    \caption{Distribution of the selection function's standard deviations (non-zero values) for the total CatNorth sample.}
    \label{fig:selection function std}
\end{figure}

In a nutshell, the selection function can be written as:
\begin{equation}
   S(\mathbf{\hat{n}})= S(A_V(\mathbf{\hat{n}}), n_\star(\mathbf{\hat{n}}),e_{G}(\mathbf{\hat{n}}), e_{P}(\mathbf{\hat{n}}),e_{W}(\mathbf{\hat{n}})),
\end{equation}
where $A_V({\mathbf{\hat{n}}})$ represents dust extinction, $n_\star(\mathbf{\hat{n}})$ denotes the stellar density, and $e_{G}(\mathbf{\hat{n}})$ refers to the detection efficiency of Gaia. 
$ e_{P}(\mathbf{\hat{n}})$, and $ e_{W}(\mathbf{\hat{n}}$) are detection efficiency of Pan-STARRS1 and CatWISE2020, respectively. We  rewrite equation \eqref{eq:model1} as:
\begin{equation}
\begin{aligned}
 n_{obs}(\mathbf{\hat{n}}) & =  S(\mathbf{\hat{n}}) n_u(\mathbf{\hat{n}})  +  S(\mathbf{\hat{n}}) n_u(\mathbf{\hat{n}}) \frac{n_\delta(\mathbf{\hat{n}})}{n_u(\mathbf{\hat{n}})}\\
 &=n_{p}(\mathbf{\hat{n}}) \left[1+\delta(\mathbf{\hat{n}})\right],
 \end{aligned}
 \label{eq:nobsnp}
\end{equation}
where $n_{p}(\mathbf{\hat{n}}) \equiv S(\mathbf{\hat{n}}) n_u(\mathbf{\hat{n}})$ represents the uniform distribution weighted by the selection function $S(\mathbf{\hat{n}})$. Here, $\delta(\mathbf{\hat{n}})$ denotes the source overdensity, which is typically modeled as a Gaussian random field. Crucially, $\delta(\mathbf{\hat{n}})$ can be treated as random noise that is statistically independent of large-scale systematic effects. Therefore we could formulate the prediction of $n_{p}(\mathbf{\hat{n}})$ as a regression problem while carefully mitigating potential overfitting that might otherwise cause the model to learn $\delta(\mathbf{\hat{n}})$. Notably, the amplitude of $\delta(\mathbf{\hat{n}})$ exhibits an inverse relationship with completeness (decreasing as the data incompleteness increases), yet this relationship does not fundamentally bias the fitting results for $n_{p}(\mathbf{\hat{n}})$. We implement this fitting using a multilayer perceptron (MLP), with the network architecture shown in Figure~\ref{fig:fnn}.

\begin{figure}
    \centering
    \includegraphics[width=1.0\linewidth]{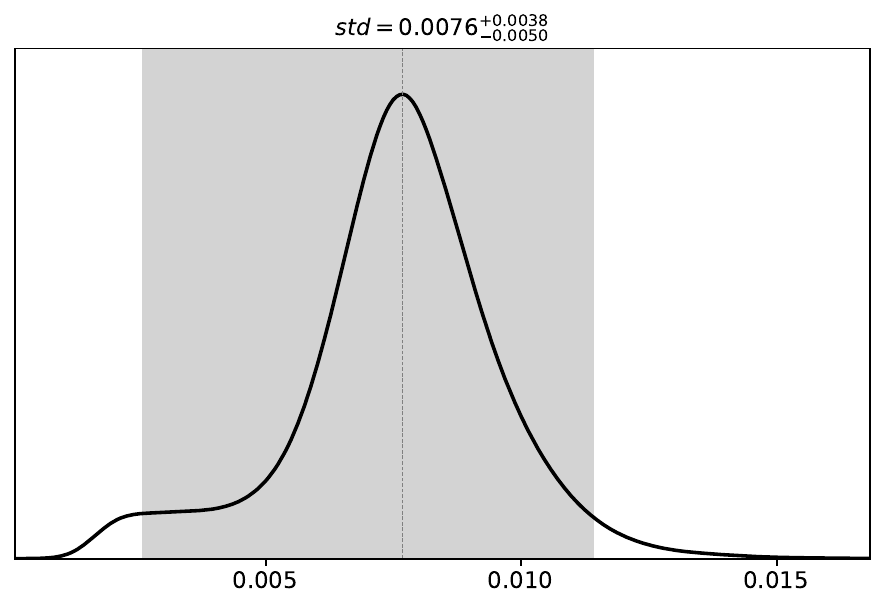}
    \caption{Probability distribution of the selection function's standard deviations (non-zero values) for the total CatNorth sample.}
    \label{fig:selection function std histgram}
\end{figure}

The selection function construction methodology proceeds through the following steps:
(\romannumeral 1) Identification of clean pixels $n_{p}^{\rm clean}(\mathbf{\hat{n}})$: Sort values in each systematics template; Select the top 15\% of pixels (e.g., choosing the smallest 15\% of values for extinction data or the largest 15\% for median magnitude maps); Define the clean pixels $n_{p}^{\rm clean}(\mathbf{\hat{n}})$ as the intersection of these selected pixels across all systematics templates. (\romannumeral 2) Normalization of $n_{p}(\mathbf{\hat{n}})$ using $\max(n_{p}^{\rm clean}(\mathbf{\hat{n}}))$, mathematically equivalent to setting $n_{u}(\mathbf{\hat{n}}) = \max(n_{p}^{\rm clean}(\mathbf{\hat{n}}))$. 
(\romannumeral 3) For pixels where the value exceeds $\max(n_{p}^{\rm clean}(\mathbf{\hat{n}}))$, we set $S(\mathbf{\hat{n}})=1$.

Figure~\ref{fig:selection function} shows the constructed $S(\mathbf{\hat{n}})$ for the complete CatNorth catalog, while Figure~\ref{fig:qso_real} presents the selection function-corrected sky density map. The comparison with Figure~\ref{fig:catnorth} demonstrates that our method successfully removes most systematic effects, producing a large-scale uniform distribution while preserving the cosmological fluctuations at smaller scales. The precision of this method primarily depends on: (1) The appropriateness of the selected systematics templates; (2) The number of systematics templates; (3) The quantity of available sources.

Compared to Gaussian process methods for selection functions \citep{2024_Storey-Fisher_quaia_catalog}, our neural network approach drastically reduces the memory usage. For $N_{\rm side}=64$ with four systematics templates, the Gaussian process requires at least 35GB of memory, while the neural network uses only $\sim$ 2GB (plus 200 MB of GPU memory). The memory usage of the Gaussian process scales quadratically with $N_{\rm side}$, whereas the neural network's memory remains stable. This efficiency enables high-resolution sky image processing and more precise selection function modeling with limited computational resources.

To test the robustness of our selection function construction method, we repeated the neural network training and selection function construction process 30 times for the full CatNorth sample. Figure \ref{fig:selection function std} displays the standard deviation map. The distribution of standard deviations across all \texttt{HEALPix} pixels is shown in Figure \ref{fig:selection function std histgram}, with a median and $1\sigma$ confidence interval of $0.0076^{+0.0038}_{-0.0050}$. This indicates that the variation in our selection function is typically less than 1$\%$.

\bibliography{sample7}{}
\bibliographystyle{aasjournalv7}

\end{CJK*}
\end{document}